\documentclass[preprint,aps,prl,superscriptaddress,longbibliography]{revtex4-1}

\usepackage{amsmath}    
\usepackage{graphicx}   
\usepackage{verbatim}   
\usepackage{color}      
\usepackage{hyperref}   
\usepackage{dcolumn}
\usepackage{amssymb}
\usepackage{bm}
\usepackage{epsfig}
\usepackage{psfrag}
\usepackage[usenames]{xcolor}
\begin{document}

\title{Synergistic action in colloidal heat engines coupled by non-conservative flows}
\author{Sudeesh Krishnamurthy}
\affiliation{Department of Physics, Indian Institute of Science, Bangalore - 560012, INDIA}
\author{Rajesh Ganapathy}
\affiliation{International Centre for Materials Science, Jawaharlal Nehru Centre for Advanced Scientific Research, Jakkur, Bangalore - 560064, INDIA}
\affiliation{Sheikh Saqr Laboratory, Jawaharlal Nehru Centre for Advanced Scientific Research, Jakkur, Bangalore - 560064, INDIA}
\author{A. K. Sood}
\affiliation{Department of Physics, Indian Institute of Science, Bangalore - 560012, INDIA}
\affiliation{International Centre for Materials Science, Jawaharlal Nehru Centre for Advanced Scientific Research, Jakkur, Bangalore - 560064, INDIA}
\date{\today}

\begin{abstract}
Collective operation of multiple engines to achieve a common objective is a vital step in the design of complex machines. Recent studies have reduced the length scales of engine design to micro and nanometers. While strategies to build complex machines from these remain to be devised, even the basic design principles remain obscure. Here, we construct and analyze the simplest collection of two engines from a pair of colloidal microspheres in optical traps at close separation. We demonstrate that at such proximity, non-conservative scattering forces that were hitherto neglected, affect the particle motion and hydrodynamics arising from dissipating these results in violating zeroth law of thermodynamics. Leveraging this in a manner analogous to microswimmers and active Brownian particles, we show that a collection of two interacting engines outperform those that are well separated. While these results explore the simplest case of two engines, the underlying concepts could aid in designing larger collections akin to biological systems.
\end{abstract}

\maketitle
\newpage

\section{Introduction}
The development of micrometer sized heat engines \cite{Bechinger_Nature,Raul_Rica,active_engine,niloy,Single_atom_engine,Szilard_engine,Lutz_Ion_engine,Lutz_Optical_engine} over the past decade has enabled the reinterpretation of laws of classical thermodynamics in a fluctuation dominated regime. A paradigmatic system used in such realizations consists of a colloidal microsphere - the working substance, in a focused laser beam that confines the particle like a piston. Thermodynamic cycles are executed by synchronized variations in intensity of the beam and temperature of the suspending medium. Engine realizations hitherto \cite{Bechinger_Nature,Raul_Rica} assumed that the optical forces exerted on the microsphere are conservative and the surrounding medium acts as a thermal bath. Experiments however have revealed that non-conservative forces are also exerted on the colloidal particle due to light scattering at the microsphere-water interface \cite{Grier_prl,Grier_vortices,Ashkin1,Sukhov_review,Florin_prl}. The resulting circulations in the probability flux termed - Brownian vortexes - could locally drive the bath out of equilibrium. Nevertheless, under isotropic conditions, where the trapped microsphere is sufficiently isolated from other particles and is at a significant distance from the boundaries of the system \cite{Volpe_vortices_negligible}, deviations due to such effects are negligible. Experimental studies in the past have also investigated the origins of the deviations when such symmetries are violated, in particular, due to trapping close to an interface \cite{Manas_vortexes}. Under such conditions, the hydrodynamic flows due to the Brownian vortexes were restricted by the presence of a surface and in turn affected the particle motion. Alternately, a surface that constrains the vortexes could also be created by trapping another colloidal particle in close proximity - a situation encountered when constructing collections of engines with multiple microspheres \cite{Hernandez_engine_array}. The ensuing vortexes in such a case would bring about non-trivial couplings and lead to co-operative behavior between the particles. Inducing hydrodynamic interactions by affecting the non-conservative flow fields - a hallmark feature of colonies of microswimmers and active Brownian particles \cite{Microswimmer1,Microswimmer2,Microswimmer4} could then be emulated in collections of optically trapped microspheres. Although experimental techniques to realize such collections of engines by generating multiple optical traps are well developed \cite{dual_trap,Grier_holo_tweezer1,Grier_holo_tweezer2}, the extent to which the non-conservative forces affect their performance remains to be quantified.

Here, we observed the operation of two heat engines, the simplest case of collective behavior and quantified the effects of non-conservative forces as they are brought close to each other. Our system consists of two colloidal microspheres trapped in adjacent optical harmonic potentials created by a pair of tightly focused laser beams as shown in Fig. \ref{Fig1}a. The optical potentials L1 and L2 were formed by focusing laser beams with mutually perpendicular polarizations at a distance $d$ from each other, in which, colloidal microspheres of diameter $\sigma$ are trapped. The intensity gradient created by a focused laser is known to apply conservative forces \cite{Ashkin} on dielectric particles with refractive index different from the surrounding medium and forms the most dominant force in the system. For the Gaussian beams used in our experiment, these forces are harmonic as illustrated in the Fig. \ref{Fig1}a. Nevertheless, the same refractive index difference also results in laser scattering from the microsphere-solution interface, and as a consequence, in a finite non-conservative force \cite{Ashkin1,Sukhov_review,Florin_prl} along the direction of propagation. The optical energy input to the bead by the scattering force is dissipated by viscous damping into the thermal energy of the surroundings. In steady state, these forces result in probabilistic circulatory motion of the particle called Brownian vortexes \cite{Grier_prl,Grier_vortices}. As noted earlier, under isotropic conditions, despite the correlated dynamics due to the ensuing hydrodynamic flows, the feedback of the energy into particle motion has been estimated to be negligible \cite{Volpe_vortices_negligible}. However, large scale particle motions could ensue if they are hindered by an interface \cite{Manas_vortexes}. In our experiments, the colloidal particle in the adjacent trap provides such an interface, albeit a fluctuating one and we exploit this to bring about a non-trivial coupling between the two particles.

\section{Results}
\subsection{Violation of zeroth law of thermodynamics}
\begin{figure*}[tbhp]
\centering
\includegraphics[width=.99\textwidth]{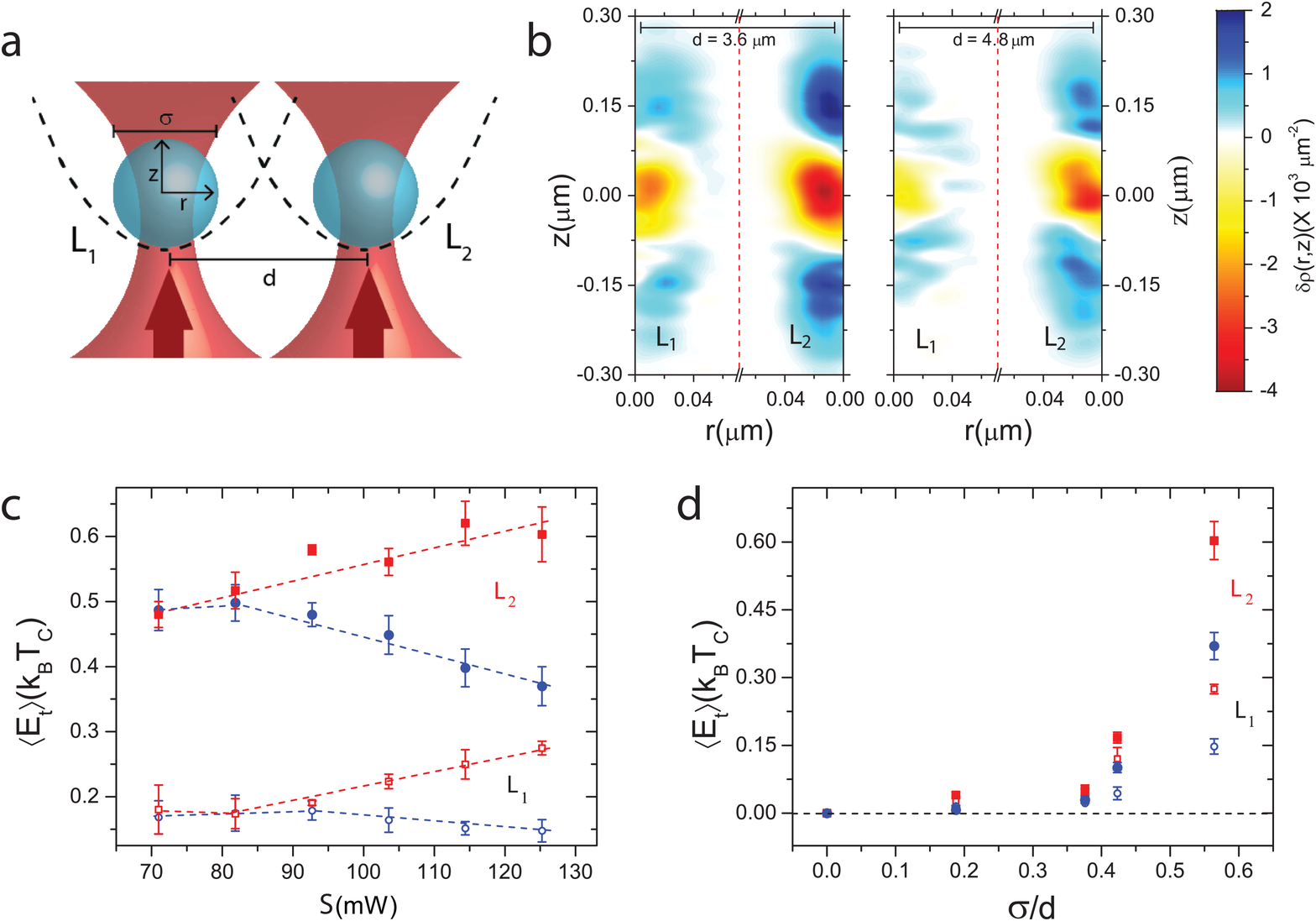}
\caption{\textbf{Violation of zeroth law of thermodynamics. a} shows schematic of experimental setup. \textbf{b} denotes change in the probability density $\delta\rho(r,z)$(represented by the scheme shown in the colorbar) on introduction of an identical bead in the adjacent trap for $d = 3.6\mu m$ and $4.8\mu m$. The total laser power $S = 125 mW$ and $S(L2)/S(L1) = 1.33$ and the solution temperature was maintained at 313K. The distributions were obtained by binning $\approx 20,000$ particle positions over boxes of size $5\mu m X 10\mu m$. The resulting distributions were further linearly interpolated onto boxes of size $0.5\mu m X 0.5\mu m$ to obtain smoother profile. \textbf{c} shows $E_t$, measured in units of $k_BT_C$, where $k_B$ is the Boltzmann constant and $T_C = 290K$, for different total laser power, $S(L1+L2)$ with $d = 3.6\mu m$ at two temperatures $T_H = 313K$(red squares) and $T_C = 290K$(blue circles) for both the traps L1 (open symbols) and L2 (closed symbols). $S(L2)/S(L1) = 1.33$ was maintained constant and the dotted lines are a guide to the eye. \textbf{d} $E_t$, at various $\sigma/d$ for the maximum total laser power $S_{max}= 125 mW$ and $\sigma = 2.03 \mu m$. $E_t$ at two temperatures $T_H$ (red squares) and $T_C$ (blue circles) are plotted for both the traps L1 (open symbols) and L2 (closed symbols). The error bars represent standard error of mean over three independent realizations}
\label{Fig1}
\end{figure*}

The central requirement to our scheme of utilizing such couplings to cooperatively extract work is that they should drive the system out of equilibrium. In the absence of such a driving, according to zeroth law of thermodynamics, two engines in contact with the same thermal reservoir are also in equilibrium with each other and average heat exchange between them is zero. Performance of such engines would remain independent of their separation. Here, we first demonstrate that restricting vortex flows results in violation of the zeroth law and allows us to exploit collective behavior. To this extent, we observed the probability distributions of particle motion in L1 and L2 at close proximity. Fig. \ref{Fig1}b shows change in the probability density of particle positions, $\delta\rho(r,z)$ of a colloidal bead trapped in L1 when another identical microsphere is introduced in L2 and vice-verse for two distances of separation $d = 3.6\mu m$ and $4.8\mu m$. $\delta\rho(r,z)$ is projected on the r-z plane of the cylindrical co-ordinates, $\textbf{r}=(r,\phi,z)$ with the laser propagation direction as the $+z$ axis. The particle positions were measured relative to the mean position of the bead in the isolated trap. The zeroth law of thermodynamics would be true only if $\delta\rho(r,z) = 0$ in the absence of potential interactions. But, from Fig. \ref{Fig1}b, $\rho(r,z)$ spreads to larger values of $r$ and $z$ on mutual interaction, with majority of the increase along the direction of laser propagation, the z-axis, where the trap stiffness is also the least. The change in the mean position of $\rho(r,z)$, however, was $\approx 5-10nm$ and negligible within the limits of the tracking resolution. Since $\delta\rho(r,z) \neq 0$, we investigated whether this could be caused by any potential interactions. Since the optical fields L1 and L2 remain constant during the observations in Fig. \ref{Fig1}b, $\delta\rho(r,z)$ is necessarily due to inter-particle interaction. The screening length of the electrostatic potential of the colloidal particle in our suspending medium, 10mM NaCl calculated using Gouy-Chapman theory was $\approx 3nm$ \cite{debye_length}, while the least surface separation between the beads used in our experiments was $1.6\mu m$. Since the counter-ions in the suspending salt solution tightly screen surface charges on the colloid, the change in $\rho(r,z)$, cannot occur due to the electrostatic double layer interactions. Further, although the motion of trapped colloidal particles is anti-correlated \cite{Quake_coupling} (See supplementary Fig. S1), hydrodynamic transfer of thermal energy between the two traps in equilibrium reservoirs is prohibited by exchange fluctuation theorem \cite{Ciliberto_prl}. Also, the increase in temperature due to laser absorption was estimated to be $\approx 0.1-0.2K$ \cite{Schmidt_heating}. Thus, change in $\rho(r,z)$, cannot be due to thermal convection currents. Also, Casimir forces between the two microspheres operate in length scales $\approx 100nm$ \cite{Casimir} and do not affect $\rho(r,z)$. Thus, the source of change in $\rho(r,z)$ is due to the interplay between the probabilistic flows generated by the trapped particles and results in a violation of the zeroth law.

To demonstrate that the probabilistic flows are sufficient to generate the interactions, we observed the changes in $\delta\rho(r,z)$ with variations in bath temperature, $T$ which mainly results in significant changes in viscosity of the solution, $\mu$ while the refractive index, $n$ remains constant. This in turn affects only the flow mediated interactions such as those generated by Brownian vortexes. We quantified the increase in $\rho(r,z)$ by evaluating the heat transferred by the reservoir through conservative optical forces, $E_t$ by integrating the corresponding potential energy along z-direction over the change in the probability density as 
\begin{equation}
E_t = \frac{1}{2}\int{k_z\delta\rho(r,z)z^2rdrdz}
\label{Et}
\end{equation}
where, $k_z$ is the trap stiffness along the z-axis. Since maximum change in $\rho(r,z)$ occurred along the direction of propagation, potential energy change in the x-y plane was found to be very small $\approx 5-10\%$ of that along z-axis and was neglected in our calculations. In Fig. \ref{Fig1}c, we measure $E_t$ at two temperatures $T_H = 313K$ and $T_C = 290K$ as total incident laser power $S = S(L1)+ S(L2)$ was varied, while the ratio of the intensities $S(L2)/S(L1) = 1.33$ was maintained constant. Across these temperatures, while the viscosity of the medium, $\mu$ changed by 1.7 times from $\mu(T_H) = 0.65 mPa\cdot s$ to $\mu(T_C) = 1.08 mPa\cdot s$ \cite{water_viscosity}, its refractive index, $n$ remained almost constant at $n(T_C) = 1.324$ and $n(T_H) = 1.322$ \cite{water_RI}. Thus, across the two temperatures, the dynamical properties of the medium were different, but the optical forces experienced by the bead remained the same. The variation of $E_t$ with $S$ in both L1 and L2 was observed to be significantly different across the two temperatures (Fig. \ref{Fig1}c) and verifies that the probabilistic flows are sufficient to produce the interactions. The trends in $E_t$ can be intuitively anticipated from the asymptotic behavior of the vortexes with $S$. At any given $T$, for a sufficiently large $S$, the conservative harmonic potential generates the most dominant force in the system and all vortexes should cease to exist. As a consequence, $E_t$ should decrease with $S$. Alternately, at low $k$, where the particle explores a significant volume, increasing $S$ results in more scattering and larger swirls that enhance $E_t$. The crossover between these behaviors is dictated by parameters such as viscosity, refractive index. Similar crossovers in particle dynamics have also been observed in previous studies \cite{Grier_vortices, Grier_minimal_model} on isolated optical traps with $k_z$ in the same range as those used in our experiment. But the $k_z$ at which they occur is harder to predict due to the chaotic nature of the vortexes. As observed in our experiment, due to lower viscosity, circulation lies below the crossover at $T_H$ and above for $T_C$. Thus, temperature essentially acts as a switch for viscosity with which synergistic interactions due to Brownian vortexes can be modulated. Further, to demonstrate that these flows are necessary for the generation of the couplings, we measure the changes in $E_t$ with $d$, along which the vortexes are known to decay. As can be seen from Fig. \ref{Fig1}d, $E_t$ decreases sharply as $d$ increases beyond 2-3 particle diameters at both $T_H$ and $T_C$. Thus, the probabilistic flows are necessary and sufficient to produce the interactions.

\begin{figure}[t]
\centering
\includegraphics[width=0.8\linewidth]{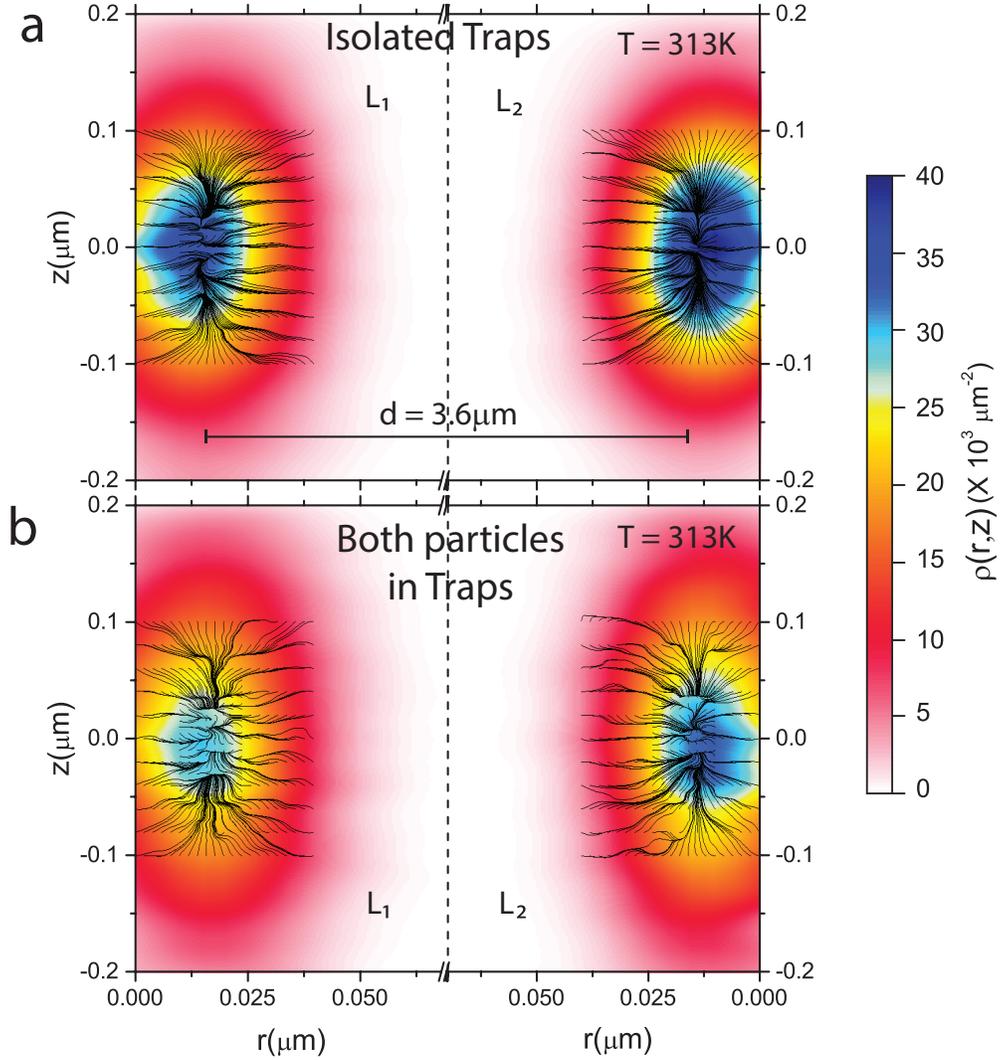}
\caption{\textbf{Vizualization of non-conservative flows} \textbf{a} shows streamlines of average particle displacement plotted in the r-z plane for both L1 and L2 for an isolated trap at $T = 313K$ with $S = S_{max}= 125 mW$ and $S(L2)/S(L1) = 1.33$. The background image represents the probability density of the particle, $\rho(r,z)$ color coded as shown in the color bar. \textbf{b} represents the same on addition of an identical bead in the adjacent trap at $d = 3.6\mu m$.}
\label{Fig2}
\end{figure}

To further obtain a visual confirmation on the central role of the underlying vortexes in generating interactions, we observed their influence on the motion of the trapped particle. In Fig. \ref{Fig2}, we plot the streamlines of velocity field of the trapped microsphere in the r-z plane at $S = 125mW$ and $T = 313K$, where maximum $E_t$ was observed in our experiments. Displacements of the particle calculated over equal intervals of time $\approx 2ms$ were spatially averaged over boxes of size $5nm \times 10nm$ to form an average velocity field and plotted as streamlines in Fig. \ref{Fig2}. Such an averaging provides us a glimpse of the particle motion in the absence of thermal noise. Intuitively, in such a scenario, particles released in the extremities should trace the streamlines and converge to the center of the trap at $(r,z)=(0,0)$, a stable fixed point had there been only conservative forces. In the presence of non-conservative forces and the associated hydrodynamic flows, however, similar to the findings in earlier literature \cite{Grier_vortices}, we observed that they converged to a region close to the $z=0$ plane, but at finite r, even in isolated optical traps as shown in Fig. \ref{Fig2}a. Streamlines in this region develop into circulatory vortex patterns akin to strange attractors due to the chaotic hydrodynamic flows. In Fig. \ref{Fig4}b, we plot the resultant streamlines on addition of a particle to the adjacent trap. We observe that the microspheres released from extremities could remain significantly farther from the z=0 plane and small changes in their initial position could result in very distinct trajectories. Thus, introducing a trapped particle in close proximity results in an increase in chaotic hydrodynamic flows that lead to energy addition by mutual interaction as anticipated by our observations. The difference in $E_t$ between $T = 313K$ and $T = 290K$ at $S=125mW$ can also be similarly expected by comparing the circulatory patterns before inserting a particle to the neighboring trap (See Supplementary Fig. S2).

\subsection{Modeling interactions and heat flows}
\begin{figure}[t]
\centering
\includegraphics[width=0.8\linewidth]{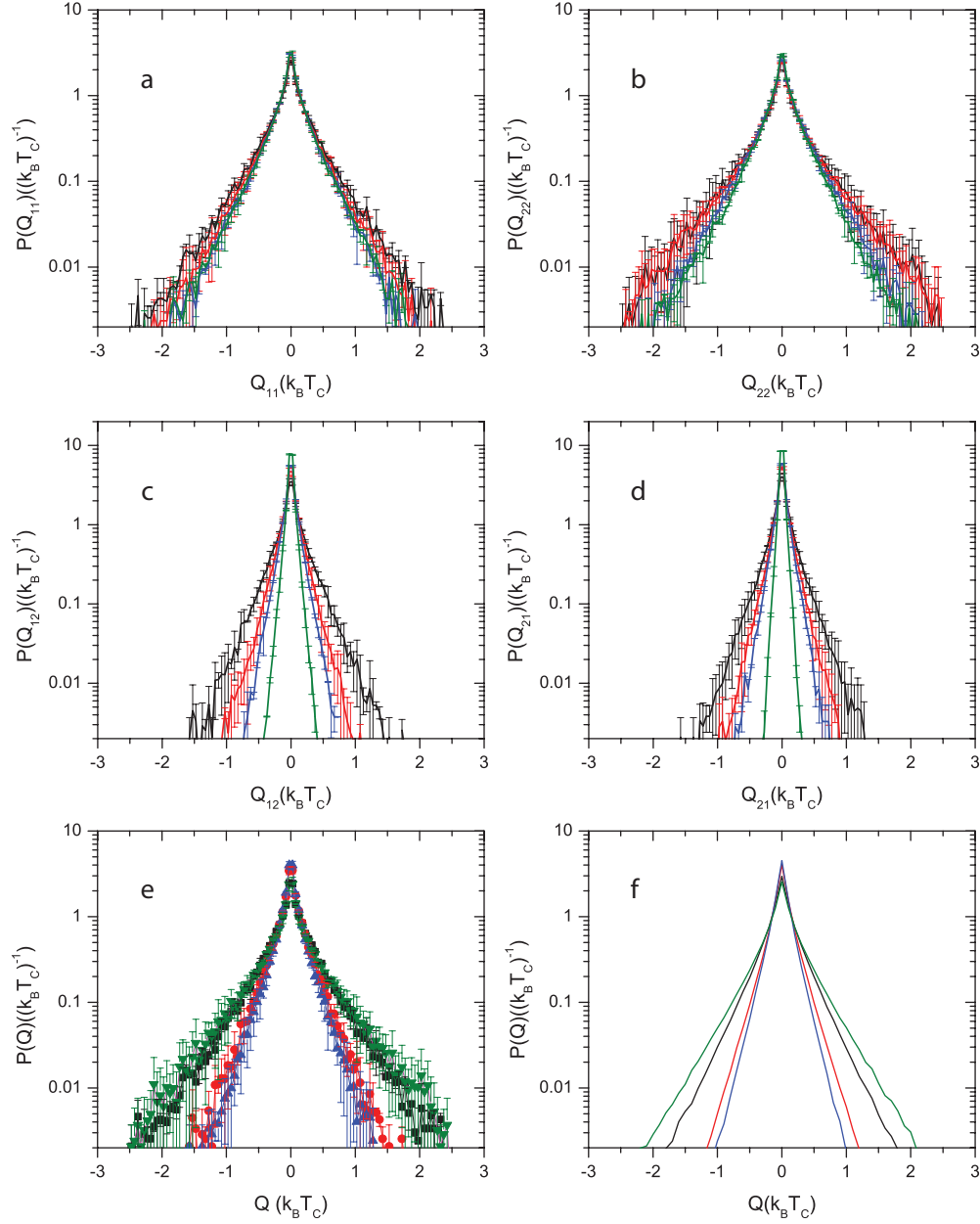}
\caption{\textbf{Heat fluxes between optically trapped particles.} \textbf{a} and \textbf{b} show probability distributions of $Q_{11}$ and $Q_{22}$, the heat fluxes due to conservative forces at various distances of separation $d = 3.6\mu m$(black), $4.8\mu m$(red), $5.4\mu m$(blue) and $10.8\mu m$(green) for the laser intensity $S = S_{max}$. \textbf{c} and \textbf{d} are similar plots of $Q_{12}$ and $Q_{21}$, the fluxes due to non-conservative hydrodynamic coupling. \textbf{e} shows probability distributions of experimentally measured heat fluxes $Q_{11}$ (black), $Q_{22}$ (green), $Q_{12}$ (red) and $Q_{21}$ (blue) for the closest separation $d = 3.6\mu m$ at the highest laser power $S = S_{max}$. \textbf{f} shows similar plots for the simulated trajectories for $\epsilon = 0.5$, $\left| \alpha_1\right| = 10\sqrt{2\gamma k_{B}T_C}$ and $\langle\left| \alpha_2\right| \rangle /\langle\left| \alpha_1\right| \rangle = 1.33$}
\label{Fig3}
\end{figure}

\begin{figure*}[t]
\centering
\includegraphics[width=0.99\linewidth]{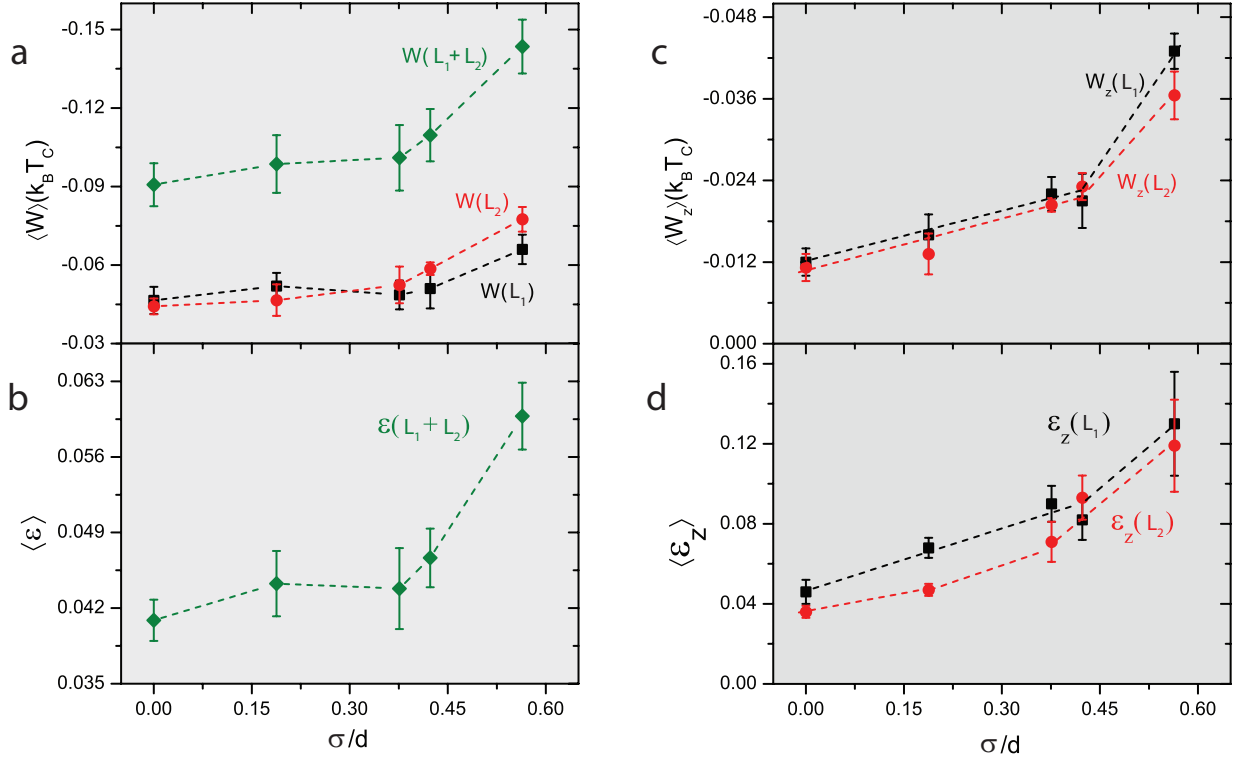}
\caption{\textbf{Performance of the engine on coupling} \textbf{a} shows $\langle W(L1 +L2)\rangle$ (green diamonds), $\langle W(L1)\rangle$ (Black squares) and $\langle W(L2)\rangle$ (Red circles) in units of $k_BT_C$ for various separation distances $d$ rescaled by $\sigma$. \textbf{b} demonstrates $\langle \epsilon(L1 +L2)\rangle$ (green diamonds) for the same. The dotted lines are a guide to eye drawn to capture the change in the trend as L1 and L2 are brought closer than 2-3 particle diameters. \textbf{c} and \textbf{d} represent similar contributions from the engine along the z-direction. The error bars represent standard error of mean over $\approx 150$ cycles.}
\label{Fig4}
\end{figure*}

Determining thermodynamic quantities to elucidate the energy flows resulting from the vortexes requires us to model the forces arising from them. Studies in the past \cite{Grier_vortices, Grier_minimal_model} have attempted to accomplish this in isolated trapped particles by employing miniature intensity profiles of incident laser beam and enumerated the associated hydrodynamics. While such models have captured general features of the vortexes, experimentally observed velocity patterns were significantly chaotic in comparison with their simulated counterparts \cite{Grier_vortices}. Our observations in Fig. \ref{Fig2} established that such behavior gets further amplified in the presence of a particle in close proximity. This motivated us to instead consider the non-conservative forces as an active noise, $\alpha_i(t)$. The magnitude of $\alpha_i(t)$ would depend on the scattering energy input to the particle which in turn depends on its position along the x-y plane. Since the position auto-correlation of the particles was approximately exponential (See supplementary information), we assumed $\langle\alpha_i(t)\alpha_i(t')\rangle \approx e^{-\frac{t-t'}{t_{\alpha}}}$. The Langevin equations of the particles is
\begin{equation}
\begin{split}
\gamma \dot{z_1} = -k_{1z} z_1 -\epsilon k_{2z} z_2 + \alpha_1(t) +\xi_1(t)\\
\gamma \dot{z_2} = -k_{2z} z_2 -\epsilon k_{1z} z_1 + \alpha_2(t) +\xi_2(t)
\end{split}
\label{eq_motion1}
\end{equation}
where $k_{iz}$ are the trap stiffness, $\gamma$ the friction constant, $\xi_i(t)$ the thermal noise and $\epsilon = \frac{3\sigma}{4d} + \frac{\sigma^3}{2d^3}$ is the hydrodynamic coupling constant calculated from the Rotne-Prager diffusion tensor \cite{Ciliberto_prl, Ciliberto_pre, herrera_jpc} (See supplementary information for detailed derivations and discussions). Heat exchanges from the particle, $Q_i$ over an interval $\tau$ can be calculated using the framework of stochastic thermodynamics \cite{Sekimoto, Seifert_review, Ciliberto_prl} as
\begin{equation}
\begin{split}
Q_i = Q_{ii} +Q_{ij}; i\neq j\\
Q_{ii} = -\int_{0}^{\tau}{k_{iz} z_i\dot{z_i}dt}\\
Q_{ij} = -\int_{0}^{\tau}{\epsilon k_{jz} z_j\dot{z_i}dt}
\end{split}
\label{heat}
\end{equation}
where, $Q_{ii}$s are the heat transferred to the reservoir by conservative forces and $Q_{ij}$s are those due to non-conservative hydrodynamic couplings. In Fig. \ref{Fig3} a-d, we plot the probability distributions of these calculated over the time step of our measurements $= 2ms$ at $S = S_{max}$ for various $d$. The variance of $Q_{11}$ and $Q_{22}$ decrease with $d$ indicating heat transfer from the reservoir at close separation (Fig.s \ref{Fig3} a $\&$ b). The variance of $Q_{12}$ and $Q_{21}$ that are known to arise due to direct hydrodynamic coupling of particle motion reduce more significantly with $d$ as anticipated (Fig.s \ref{Fig3} c $\&$ d), but $\langle Q_{12}\rangle = 0$ and $\langle Q_{21}\rangle = 0$ indicating the absence of net heat transfer between the particles. Thus, while the energy dumped into the reservoir by vortex flows is being reused by inducing additional fluctuations into particle motion, mean heat exchanged between the particles through hydrodynamic coupling is negligible in the steady state.

Although the analysis in Fig. \ref{Fig3} a-d presents a comprehensive picture of the underlying heat flows, Eqn.s \ref{heat} do not explicitly require nature of the noise as an input. To check if the active noise observed in these measurements indeed match with our intuition, particularly on the correlation $\langle\alpha_i(t)\alpha_i(t')\rangle \approx e^{-\frac{t-t'}{t_{\alpha}}}$, we followed the motion of the particle due to eqn. \ref{eq_motion1} through Brownian dynamics simulation. We assumed $t_{\alpha} = 10ms$, the approximate decay time of position correlations along x and y axis and maintained $\langle \left|\alpha_2\right| \rangle/\langle\left|\alpha_1\right|\rangle = 1.33$ to reflect the ratio of intensities. A comparison between the heat flows calculated from the resulting trajectories for $\left|\alpha_1\right| = 10 \sqrt{2\gamma k_B T_C}$ (Fig. \ref{Fig3}f) with the experimental data for $S = S_{max}$ and $d = 3.6\mu m$ (Fig. \ref{Fig3}e) shows a good agreement between them. Thus, our model for the forces described by the equations of motion Eqn.s \ref{eq_motion1} capture the energy flows well and can be used to analyze performance of engines coupled by vortexes. Unlike experiments however, the simulations allowed us to independently vary the magnitude of $\alpha_i$ and $\epsilon$ and delineate the contributions to heat transfer from enhancement of vortexes and hydrodynamic coupling. We observed that the distributions of heat flows obtained by varying $\left|\alpha_i\right|$ by maintaining $\epsilon$ constant matched closely with those in Fig. \ref{Fig3} a-d (See supplementary figure Fig. S5 for more information). The viceverse, however resulted in negligible changes (See supplementary figure Fig. S6 for more information). This indicates that while hydrodynamic coupling does influence the heat transferred to the particle, majority of the contribution arises from enhancement of vortex flows. While the analysis in Fig. \ref{Fig3} mainly compared simulations with the interactions at $S = S_{max}$, we do explore a range of laser intensities and solution temperatures in our experiments like in Fig. \ref{Fig2}. The heat flows in such a situation could be arrived at by setting an appropriate value for $\left|\alpha_1\right|$ (See supplementary figure Fig. S7a for more information). Also, we had empirically fixed $\langle\left| \alpha_2\right| \rangle/\langle\left| \alpha_1\right| \rangle = 1.33$ based only on the ratio of laser intensities and not considering the associated hydrodynamics. Varying this could allow us to further adjust our model to capture experimental behavior (See supplementary figure Fig. S7b for more information).

In a more general sense, the interaction between the colloidal particles caused by restricting vortex flows is analogous to collections of microswimmers and active Brownian particles(ABP) \cite{Microswimmer1, Microswimmer2, Microswimmer4} at close proximity. Eqn. \ref{eq_motion1} could equivalently be used to describe ABP s in harmonic potentials in close separation. Intuitively, a typical ABP draws energy from a driving source through an anisotropic forcing comprising of chemical, thermal etc. interactions and dissipates it into hydrodynamics of the surrounding medium, thus propelling it in a select direction \cite{Bechinger_rmp}. Similarly, a colloidal particle in an optical trap receives energy along the direction of laser propagation from non-conservative forces through laser scattering and dissipates it into the Brownian vortexes. However, the particle is also under the influence of a strong harmonic potential and laser scattering only shifts its mean position in the optical trap. At close separation, the hydrodynamics created by the ABP s while dissipating the energy received from the active driving force is known to influence the motion an adjacent particle. Our observations in Fig.s \ref{Fig1}-\ref{Fig3} presents a similar scenario with trapped colloidal particles which results in violation of zeroth law of thermodynamics. This enabled us to construct engine cycles that exploit the collective behavior.

\subsection{Heat engines coupled by Brownian vortexes}
\begin{figure}[!t]
\centering
\includegraphics[width=0.4\linewidth]{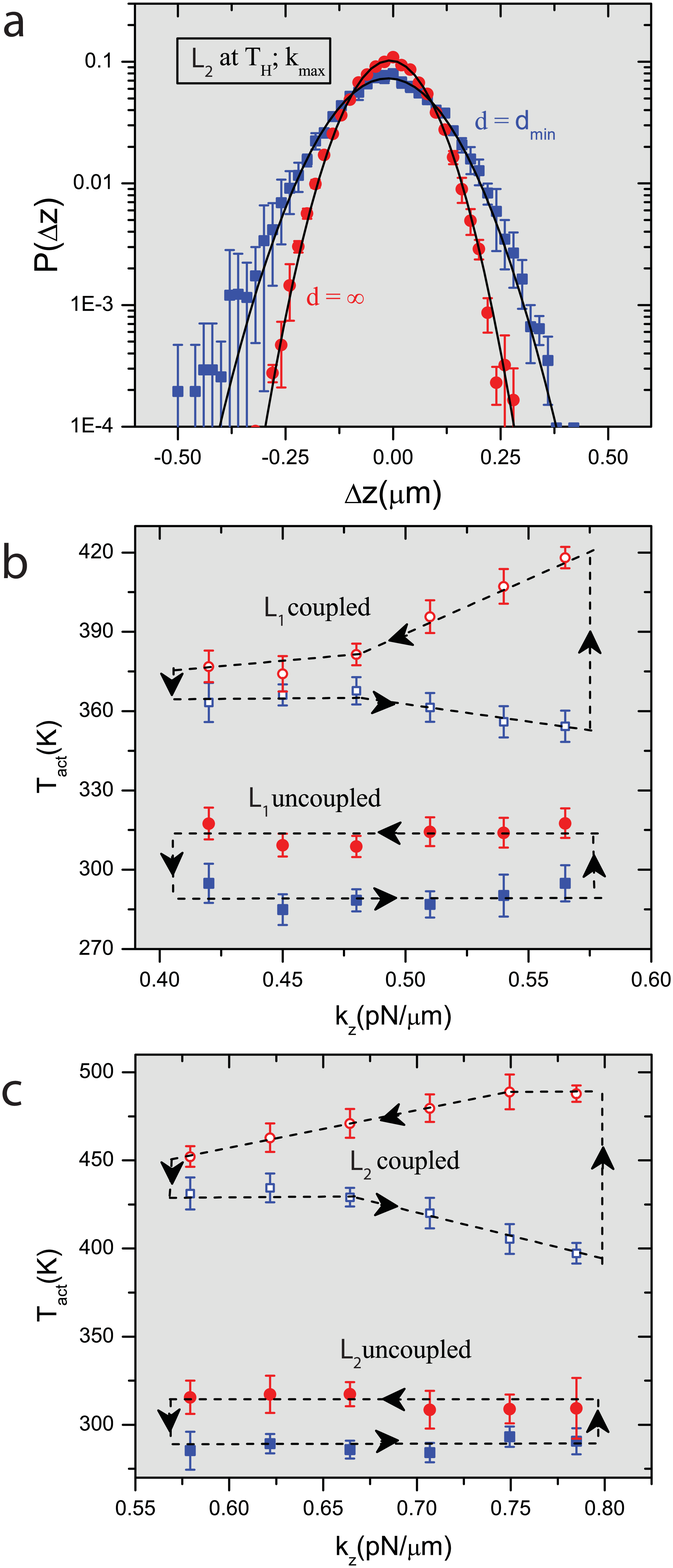}
\caption{\textbf{Elucidating the origins of superior performance. a} $P(\Delta z)$ of a particle in L2 before(red squares) and after(blue circles) addition of another colloidal bead in L1 for $T = T_H$, $d = d_{min} = 3.6\mu m$ and $k = k_{max}$, where, the largest $E_t$ was observed in Fig 3, is plotted. The solid lines represent Gaussian fits for the same. \textbf{b} and \textbf{c} show the trajectory of the system in the $k_z-T_{eff}$ plane for the engine in L1 and L2 respectively as Stirling cycle is performed before(white filled symbols) and after(color filled symbols) introducing an identical particle in the adjacent trap. In the above experiments $d = d_{min} = 3.6 \mu m$, $T_H = 313K$(Red circles) and $T_C = 290K$(Blue squares). The error bars are standard error of mean over three independent realizations.}
\label{Fig5}
\end{figure}

With colloidal particles in both the optical traps L1 and L2, we executed the microscopic equivalent of Stirling cycle \cite{Bechinger_Nature,Raul_Rica,active_engine} (See supplementary movie) for various distances of separation. The isothermal processes were performed by changing the total laser power, $S$ between $(S_{max},S_{min}) = (71mW,125mW)$ as the temperature of the suspending medium was maintained constant. The isochoric processes were realized by passing heat exchanging fluid in an adjacent channel \cite{active_engine}, which modulates the temperature between $(T_H,T_C) = (313K,290K)$ as $S$ is maintained constant. The engines are operated synchronously without any phase difference with a cycle time of $22s$ (7s for each isotherm and 4s for each isochore). As relaxation times for hydrodynamic transfer of energy between the microspheres is $\approx 10ms$, the engines are effectively operated in the quasi-static limit. Since the engine bath interactions at small $d$ would now include the mutual interaction between the flows generated by Brownian vortexes, which depend on $S$, the effective protocol by the engine deviates, particularly along the isotherms from the Stirling cycle. Nevertheless, to compare the performance across various $d$, we followed the same protocol at all $d$ as for $d = \infty$. The performance of the engine in these realizations at various $d$ (rescaled by $\sigma$) are plotted in Fig. \ref{Fig5}. The work done and efficiency were calculated by tracing the particle trajectories and using the framework of stochastic thermodynamics \cite{active_engine,Sekimoto, Seifert_review} generated for our model in Eqn. \ref{eq_motion1} (See supplementary information for detailed derivations and discussions). Due to the variation of $S$ in the isothermal cycles, the stiffness of this potential varied between $(k_{z min},k_{z max}) = (0.41\pm 0.02,0.58\pm 0.01) pN/\mu m$ for L1 and $(k_{z min},k_{z max}) = (0.56\pm 0.01,0.795\pm 0.02) pN/\mu m$ for L2. From the Fig. \ref{Fig5} a, we see that the average total work done by the engine, $\langle W(L1 +L2)\rangle$ measured from trajectories along all the three axes x,y and z, in units of $k_BT_C$ increase sharply along with the individual work done by both the traps $\langle W(L1)\rangle$ and $\langle W(L2)\rangle$, as $\sigma/d \rightarrow 1$. As a result of this, the average total efficiency $\epsilon(L1+L2)$ plotted in Fig. \ref{Fig5} b also increases simultaneously. Most of this enhancement in performance, however, occurs when L1 and L2 are closer than 2-3 particle diameters, as can be seen from the dotted lines drawn to capture this trend.

To isolate the contribution from the changes in vortexes, we plot the average work done along the z-direction, $\langle W_z\rangle$ measured in units of $k_BT_C$ for the individual traps in Fig. \ref{Fig5}c. As can be seen from the figure, $\langle W_z\rangle$ increases by 3.5 times as the engines are brought into close proximity. About $70\%$ of this occurs at separation distances in the range of 2-3 particle diameters, where, as described in Fig. \ref{Fig1}d energy is input into the particle motion from the vortexes. Thus, most of the increase in $\langle W(L1 +L2)\rangle$ can be attributed to the increase in $\langle W_z\rangle$. As a result of this, the efficiency calculated along the z-axis, $\epsilon_z$ for both L1 and L2 (Fig. \ref{Fig5}d) increase drastically by about 3 times and the increase in $\epsilon(L1+L2)$ can be attributed to the better performance of the engine along the z-axis.

To elucidate the origins of increase in the performance of the engines due to operation at close proximity, we trace the trajectory of the system in the state space of the Stirling cycle - the k-T plane. However, an effective temperature that describes the fluctuations in the system can be defined only when the probability distribution of the particle position, $P(\Delta z)$ resemble those due to a thermal reservoir, i.e. follow a Gaussian profile for our harmonic optical trap \cite{Bocquet}. A system with an additional non-equilibrium noise, as in our experiment, follows such a distribution only if either the potential is sufficiently wide, or amplitude of the active fluctuations are significantly small. To verify if even one of this holds true in our case, we plot $P(\Delta z)$ in Fig. \ref{Fig5} a for L2 at $k_{max}$ with a bath temperature at $T_H$, where, maximum $E_t$ was detected in our setup (Fig. \ref{Fig1}c). Deviations from an `Equilibrium-like' behavior, if any, should show up under these extreme conditions. As established earlier (Fig. \ref{Fig1}b), we observe that the $P(\Delta z)$ widens on interaction (Fig. \ref{Fig5} a). The solid lines which represent Gaussian fits to $P(\Delta z)$ match well with the experimental measurements only in the first decade but deviate from the fit in the tails. The increase in the energy encompassed by the Gaussian fits measured by averaging potential energy across the distribution \cite{active_engine} was found to be about $\approx 0.41 k_BT_C$. This accounts for only $72\%$ of the total $E_t$ associated with the entire $P(\Delta z)$. Thus, the synergistic action simultaneously affects both, the width and the statistics of the system. This is in contrast with observations in earlier literature \cite{Grier_prl} in isolated traps, where, $P(\Delta z)$ was shown to be Gaussian. Thus, the coupling of the vortexes causes a significant input of energy into the particle motion. Nevertheless, a $T_{act}$ derived from the potential energy of the system can still be defined by following protocols in previous studies on active engines \cite{active_engine}, using equipartition theorem $\frac{1}{2}k_B T_{act} = \frac{1}{2}k_z \langle z^2\rangle$ (See supplementary information for discussions). As in the case of $E_t$, we consider motion only along the z-direction to calculate $T_{act}$, since the addition of another particle in the adjacent trap causes negligible changes in $P(\Delta x)$ and $P(\Delta y)$ (Fig. \ref{Fig1}b, also see Supplementary Fig. S3).

We can now plot the state of the system on the $k_z-T_{act}$ plane as shown in Fig. \ref{Fig5} b and c. As anticipated from Fig. \ref{Fig1} c, the coupling between the particle motion results in deviations from the effective Stirling protocol. Since $E_t$ has a strong dependence on $S$, significant deviations were observed along the isotherms, where, $S$ is adjusted to vary the trap stiffness/confinement, $k_z$ at constant bath temperature, $T$. As a consequence, the path traced by the system is no longer a rectangle, but a distorted trapezoid. Although the area in the $k_z-T_{act}$ plane increases on coupling (Fig.s \ref{Fig5} b and c), it does not necessarily represent the work done. Nevertheless, the difference in $T_{act}$ between the hot and cold 'isotherms' is proportional to work done and increases on introducing the adjacent particle. Thus, distortions of the effective Stirling protocol lead to enhancement of the work done and efficiency.
The source of the performance, however, is not just the energy added by the neighboring trap, but the difference in the increase in $T_{act}$ between the hot and cold reservoirs, which in-turn depends crucially on the nature of the vortexes.

\section{Discussions}
In conclusion, we have exploited the non-conservative forces in focused laser beams, to bring about a non-trivial coupling between two micrometer sized colloidal heat engines. This resulted in violation of the zeroth law of thermodynamics and allowed us to induce collective behavior. The total performance of the engines increased due to the energy addition from the associated flows due to Brownian vortexes when they were operated in close proximity. Although such vortexes existed even in isolated traps, the optical energy input from the non-conservative forces was rapidly dissipated into the surroundings. Our results demonstrate that in the presence of a neighboring particle, this energy can be reused to extract work. Thus, even though the energy input remains the same, if engines that generate non-conservative flows are operated in close vicinity, they utilize the same energy intake better. While the nature of vortexes in the general case of a large collection of microspheres might be harder to analyze, our results establish that unlike isolated engines, the non-conservative couplings between them could influence the total performance. Hydrodynamic flows generated by non-conservative forces are also characteristic to all microswimmers \cite{Microswimmer1,Microswimmer2,Microswimmer4} and self-propelled active Brownian particles \cite{Bechinger_rmp} driven by electric and magnetic fields \cite{Bartolo,Magnetic_rotors}, chemical reactions \cite{Microswimmer3,Sanchez} etc. Interactions in collections of such particles are also mediated by the hydrodynamic flows \cite{Microswimmer1,Microswimmer2,Microswimmer3,Microswimmer4,Hydrodynamics2}. We demonstrated that models used to describe active Brownian particles could also capture interactions in our system. It would be tempting to consider whether the superior performance observed in our experiments could also be a reflection of the enhancement of propulsion velocity observed in collections of such microswimmers \cite{Microswimmer1,Microswimmer2,Microswimmer3,Microswimmer4,Magnetic_rotors}.

\section{Materials and methods}
The polystyrene colloidal microspheres (mean diameter, $\sigma = 2.03\mu m$) used for realization of the heat engine in our experiment were obtained from Bangslabs, USA. The particles were trapped in 10mM NaCl solution at a distance of $25\mu m$ from the surface of a coverslip by tightly focusing a NDYVO4 IR laser beam of wavelength, $\lambda_0 = 1064nm$ with a Carl Zeiss 100X objective with numerical aperture, N.A. = 1.4 mounted on a Carl Zeiss Axiovert Microscope. While trapping in a salt solution prevented double layer interactions between the particles, the sufficiently large distance from the coverslip avoided boundary effects on the system. To create two adjacent optical traps, we split the incident laser beam into perpendicularly polarized components using a polarizing beam splitter, which were then independently steered into the back aperture of the objective. Since, the vortexes created by the non-conservative forces is sensitive to parameters such as orientation of the beams, difference between their intensities, position of the focus etc., the steering process would create two distinct optical traps, despite being operated at the same power. In our experiments, however, we maintained the ratio between the intensities at $S(L2)/S(L1) = 1.33$ and studied the most general case of the coupling. The beam waist at the focal point, which indicates the region of influence of the laser was $\approx \frac{2\lambda_0}{\pi N.A.} = 480nm$ and significantly lesser than the closest separation distance between the microspheres $\approx 3.6\mu m$ used in our experiment. To mark the end points of the isochoric process and to monitor the distance from the coverslip, a very low powered red laser (Thorlabs ML101J8 Diode laser of wavelength 632 nm controlled using a Thorlabs TCLDM9 temperature controlled laser diode module) aligned at a slanting angle was shined at a distance at least $10\mu m$ from the trapped particles. The red laser was switched on only during the isochoric processes and its reflection from the bottom coverslip of the sample cell was monitored. This provided us the necessary temporal resolution to resolve the isochoric process as well as spatial resolution of $0.2\mu m$ to monitor the distance from the coverslip. The solution temperature was tuned by flowing heat exchanging fluid in an adjacent chamber used in earlier experiments on active engines \cite{active_engine}. Particles were imaged using a Basler Ace 180 kc color camera at $500 frames/sec$. The influence of red laser was eliminated by considering only the green slice of the RGB image and the particle position is tracked to sub-pixel resolution to an accuracy of 5 nm along x-y and 10 nm along z direction using custom built tracking codes in Matlab.

\section*{Acknowledgements}
AKS and SK thank the Department of Science and Technology, India for financial support under the Year of Science Professorship and Nanomission council.

\bibliography{paper_ref}

\begin{thebibliography}{42}%
\makeatletter
\providecommand \@ifxundefined [1]{%
 \@ifx{#1\undefined}
}%
\providecommand \@ifnum [1]{%
 \ifnum #1\expandafter \@firstoftwo
 \else \expandafter \@secondoftwo
 \fi
}%
\providecommand \@ifx [1]{%
 \ifx #1\expandafter \@firstoftwo
 \else \expandafter \@secondoftwo
 \fi
}%
\providecommand \natexlab [1]{#1}%
\providecommand \enquote  [1]{``#1''}%
\providecommand \bibnamefont  [1]{#1}%
\providecommand \bibfnamefont [1]{#1}%
\providecommand \citenamefont [1]{#1}%
\providecommand \href@noop [0]{\@secondoftwo}%
\providecommand \href [0]{\begingroup \@sanitize@url \@href}%
\providecommand \@href[1]{\@@startlink{#1}\@@href}%
\providecommand \@@href[1]{\endgroup#1\@@endlink}%
\providecommand \@sanitize@url [0]{\catcode `\\12\catcode `\$12\catcode
  `\&12\catcode `\#12\catcode `\^12\catcode `\_12\catcode `\%12\relax}%
\providecommand \@@startlink[1]{}%
\providecommand \@@endlink[0]{}%
\providecommand \url  [0]{\begingroup\@sanitize@url \@url }%
\providecommand \@url [1]{\endgroup\@href {#1}{\urlprefix }}%
\providecommand \urlprefix  [0]{URL }%
\providecommand \Eprint [0]{\href }%
\providecommand \doibase [0]{http://dx.doi.org/}%
\providecommand \selectlanguage [0]{\@gobble}%
\providecommand \bibinfo  [0]{\@secondoftwo}%
\providecommand \bibfield  [0]{\@secondoftwo}%
\providecommand \translation [1]{[#1]}%
\providecommand \BibitemOpen [0]{}%
\providecommand \bibitemStop [0]{}%
\providecommand \bibitemNoStop [0]{.\EOS\space}%
\providecommand \EOS [0]{\spacefactor3000\relax}%
\providecommand \BibitemShut  [1]{\csname bibitem#1\endcsname}%
\let\auto@bib@innerbib\@empty
\bibitem [{\citenamefont {Blickle}\ and\ \citenamefont
  {Bechinger}(2012)}]{Bechinger_Nature}%
  \BibitemOpen
  \bibfield  {author} {\bibinfo {author} {\bibfnamefont {Valentin}\
  \bibnamefont {Blickle}}\ and\ \bibinfo {author} {\bibfnamefont {Clemens}\
  \bibnamefont {Bechinger}},\ }\bibfield  {title} {\enquote {\bibinfo {title}
  {Realization of a micrometre-sized stochastic heat engine},}\ }\href@noop {}
  {\bibfield  {journal} {\bibinfo  {journal} {Nature Physics}\ }\textbf
  {\bibinfo {volume} {8}},\ \bibinfo {pages} {143--146} (\bibinfo {year}
  {2012})}\BibitemShut {NoStop}%
\bibitem [{\citenamefont {Mart{\'\i}nez}\ \emph {et~al.}(2016)\citenamefont
  {Mart{\'\i}nez}, \citenamefont {Rold{\'a}n}, \citenamefont {Dinis},
  \citenamefont {Petrov}, \citenamefont {Parrondo},\ and\ \citenamefont
  {Rica}}]{Raul_Rica}%
  \BibitemOpen
  \bibfield  {author} {\bibinfo {author} {\bibfnamefont {Ignacio~A}\
  \bibnamefont {Mart{\'\i}nez}}, \bibinfo {author} {\bibfnamefont {{\'E}dgar}\
  \bibnamefont {Rold{\'a}n}}, \bibinfo {author} {\bibfnamefont {Luis}\
  \bibnamefont {Dinis}}, \bibinfo {author} {\bibfnamefont {Dmitri}\
  \bibnamefont {Petrov}}, \bibinfo {author} {\bibfnamefont {Juan~MR}\
  \bibnamefont {Parrondo}}, \ and\ \bibinfo {author} {\bibfnamefont
  {Ra{\'u}l~A}\ \bibnamefont {Rica}},\ }\bibfield  {title} {\enquote {\bibinfo
  {title} {Brownian carnot engine},}\ }\href@noop {} {\bibfield  {journal}
  {\bibinfo  {journal} {Nature physics}\ }\textbf {\bibinfo {volume} {12}},\
  \bibinfo {pages} {67--70} (\bibinfo {year} {2016})}\BibitemShut {NoStop}%
\bibitem [{\citenamefont {Krishnamurthy}\ \emph {et~al.}(2016)\citenamefont
  {Krishnamurthy}, \citenamefont {Ghosh}, \citenamefont {Chatterji},
  \citenamefont {Ganapathy},\ and\ \citenamefont {Sood}}]{active_engine}%
  \BibitemOpen
  \bibfield  {author} {\bibinfo {author} {\bibfnamefont {Sudeesh}\ \bibnamefont
  {Krishnamurthy}}, \bibinfo {author} {\bibfnamefont {Subho}\ \bibnamefont
  {Ghosh}}, \bibinfo {author} {\bibfnamefont {Dipankar}\ \bibnamefont
  {Chatterji}}, \bibinfo {author} {\bibfnamefont {Rajesh}\ \bibnamefont
  {Ganapathy}}, \ and\ \bibinfo {author} {\bibfnamefont {AK}~\bibnamefont
  {Sood}},\ }\bibfield  {title} {\enquote {\bibinfo {title} {A micrometre-sized
  heat engine operating between bacterial reservoirs},}\ }\href@noop {}
  {\bibfield  {journal} {\bibinfo  {journal} {Nature Physics}\ }\textbf
  {\bibinfo {volume} {12}},\ \bibinfo {pages} {1134--1138} (\bibinfo {year}
  {2016})}\BibitemShut {NoStop}%
\bibitem [{\citenamefont {Roy}\ \emph {et~al.}(2021)\citenamefont {Roy},
  \citenamefont {Leroux}, \citenamefont {Sood},\ and\ \citenamefont
  {Ganapathy}}]{niloy}%
  \BibitemOpen
  \bibfield  {author} {\bibinfo {author} {\bibfnamefont {Niloyendu}\
  \bibnamefont {Roy}}, \bibinfo {author} {\bibfnamefont {Nathan}\ \bibnamefont
  {Leroux}}, \bibinfo {author} {\bibfnamefont {AK}~\bibnamefont {Sood}}, \ and\
  \bibinfo {author} {\bibfnamefont {Rajesh}\ \bibnamefont {Ganapathy}},\
  }\bibfield  {title} {\enquote {\bibinfo {title} {Tuning the performance of a
  micrometer-sized stirling engine through reservoir engineering},}\
  }\href@noop {} {\bibfield  {journal} {\bibinfo  {journal} {arXiv preprint
  arXiv:2101.08506}\ } (\bibinfo {year} {2021})}\BibitemShut {NoStop}%
\bibitem [{\citenamefont {Ro{\ss}nagel}\ \emph {et~al.}(2016)\citenamefont
  {Ro{\ss}nagel}, \citenamefont {Dawkins}, \citenamefont {Tolazzi},
  \citenamefont {Abah}, \citenamefont {Lutz}, \citenamefont {Schmidt-Kaler},\
  and\ \citenamefont {Singer}}]{Single_atom_engine}%
  \BibitemOpen
  \bibfield  {author} {\bibinfo {author} {\bibfnamefont {Johannes}\
  \bibnamefont {Ro{\ss}nagel}}, \bibinfo {author} {\bibfnamefont {Samuel~T}\
  \bibnamefont {Dawkins}}, \bibinfo {author} {\bibfnamefont {Karl~N}\
  \bibnamefont {Tolazzi}}, \bibinfo {author} {\bibfnamefont {Obinna}\
  \bibnamefont {Abah}}, \bibinfo {author} {\bibfnamefont {Eric}\ \bibnamefont
  {Lutz}}, \bibinfo {author} {\bibfnamefont {Ferdinand}\ \bibnamefont
  {Schmidt-Kaler}}, \ and\ \bibinfo {author} {\bibfnamefont {Kilian}\
  \bibnamefont {Singer}},\ }\bibfield  {title} {\enquote {\bibinfo {title} {A
  single-atom heat engine},}\ }\href@noop {} {\bibfield  {journal} {\bibinfo
  {journal} {Science}\ }\textbf {\bibinfo {volume} {352}},\ \bibinfo {pages}
  {325--329} (\bibinfo {year} {2016})}\BibitemShut {NoStop}%
\bibitem [{\citenamefont {Koski}\ \emph {et~al.}(2014)\citenamefont {Koski},
  \citenamefont {Maisi}, \citenamefont {Pekola},\ and\ \citenamefont
  {Averin}}]{Szilard_engine}%
  \BibitemOpen
  \bibfield  {author} {\bibinfo {author} {\bibfnamefont {Jonne~V}\ \bibnamefont
  {Koski}}, \bibinfo {author} {\bibfnamefont {Ville~F}\ \bibnamefont {Maisi}},
  \bibinfo {author} {\bibfnamefont {Jukka~P}\ \bibnamefont {Pekola}}, \ and\
  \bibinfo {author} {\bibfnamefont {Dmitri~V}\ \bibnamefont {Averin}},\
  }\bibfield  {title} {\enquote {\bibinfo {title} {Experimental realization of
  a szilard engine with a single electron},}\ }\href@noop {} {\bibfield
  {journal} {\bibinfo  {journal} {Proceedings of the National Academy of
  Sciences}\ }\textbf {\bibinfo {volume} {111}},\ \bibinfo {pages}
  {13786--13789} (\bibinfo {year} {2014})}\BibitemShut {NoStop}%
\bibitem [{\citenamefont {Abah}\ \emph {et~al.}(2012)\citenamefont {Abah},
  \citenamefont {Rossnagel}, \citenamefont {Jacob}, \citenamefont {Deffner},
  \citenamefont {Schmidt-Kaler}, \citenamefont {Singer},\ and\ \citenamefont
  {Lutz}}]{Lutz_Ion_engine}%
  \BibitemOpen
  \bibfield  {author} {\bibinfo {author} {\bibfnamefont {Obinna}\ \bibnamefont
  {Abah}}, \bibinfo {author} {\bibfnamefont {Johannes}\ \bibnamefont
  {Rossnagel}}, \bibinfo {author} {\bibfnamefont {Georg}\ \bibnamefont
  {Jacob}}, \bibinfo {author} {\bibfnamefont {Sebastian}\ \bibnamefont
  {Deffner}}, \bibinfo {author} {\bibfnamefont {Ferdinand}\ \bibnamefont
  {Schmidt-Kaler}}, \bibinfo {author} {\bibfnamefont {Kilian}\ \bibnamefont
  {Singer}}, \ and\ \bibinfo {author} {\bibfnamefont {Eric}\ \bibnamefont
  {Lutz}},\ }\bibfield  {title} {\enquote {\bibinfo {title} {Single-ion heat
  engine at maximum power},}\ }\href@noop {} {\bibfield  {journal} {\bibinfo
  {journal} {Physical review letters}\ }\textbf {\bibinfo {volume} {109}},\
  \bibinfo {pages} {203006} (\bibinfo {year} {2012})}\BibitemShut {NoStop}%
\bibitem [{\citenamefont {Dechant}\ \emph {et~al.}(2015)\citenamefont
  {Dechant}, \citenamefont {Kiesel},\ and\ \citenamefont
  {Lutz}}]{Lutz_Optical_engine}%
  \BibitemOpen
  \bibfield  {author} {\bibinfo {author} {\bibfnamefont {Andreas}\ \bibnamefont
  {Dechant}}, \bibinfo {author} {\bibfnamefont {Nikolai}\ \bibnamefont
  {Kiesel}}, \ and\ \bibinfo {author} {\bibfnamefont {Eric}\ \bibnamefont
  {Lutz}},\ }\bibfield  {title} {\enquote {\bibinfo {title} {All-optical
  nanomechanical heat engine},}\ }\href@noop {} {\bibfield  {journal} {\bibinfo
   {journal} {Physical review letters}\ }\textbf {\bibinfo {volume} {114}},\
  \bibinfo {pages} {183602} (\bibinfo {year} {2015})}\BibitemShut {NoStop}%
\bibitem [{\citenamefont {Roichman}\ \emph {et~al.}(2008)\citenamefont
  {Roichman}, \citenamefont {Sun}, \citenamefont {Stolarski},\ and\
  \citenamefont {Grier}}]{Grier_prl}%
  \BibitemOpen
  \bibfield  {author} {\bibinfo {author} {\bibfnamefont {Yohai}\ \bibnamefont
  {Roichman}}, \bibinfo {author} {\bibfnamefont {Bo}~\bibnamefont {Sun}},
  \bibinfo {author} {\bibfnamefont {Allan}\ \bibnamefont {Stolarski}}, \ and\
  \bibinfo {author} {\bibfnamefont {David~G}\ \bibnamefont {Grier}},\
  }\bibfield  {title} {\enquote {\bibinfo {title} {Influence of nonconservative
  optical forces on the dynamics of optically trapped colloidal spheres: the
  fountain of probability},}\ }\href@noop {} {\bibfield  {journal} {\bibinfo
  {journal} {Physical review letters}\ }\textbf {\bibinfo {volume} {101}},\
  \bibinfo {pages} {128301} (\bibinfo {year} {2008})}\BibitemShut {NoStop}%
\bibitem [{\citenamefont {Sun}\ \emph {et~al.}(2009)\citenamefont {Sun},
  \citenamefont {Lin}, \citenamefont {Darby}, \citenamefont {Grosberg},\ and\
  \citenamefont {Grier}}]{Grier_vortices}%
  \BibitemOpen
  \bibfield  {author} {\bibinfo {author} {\bibfnamefont {Bo}~\bibnamefont
  {Sun}}, \bibinfo {author} {\bibfnamefont {Jiayi}\ \bibnamefont {Lin}},
  \bibinfo {author} {\bibfnamefont {Ellis}\ \bibnamefont {Darby}}, \bibinfo
  {author} {\bibfnamefont {Alexander~Y}\ \bibnamefont {Grosberg}}, \ and\
  \bibinfo {author} {\bibfnamefont {David~G}\ \bibnamefont {Grier}},\
  }\bibfield  {title} {\enquote {\bibinfo {title} {Brownian vortexes},}\
  }\href@noop {} {\bibfield  {journal} {\bibinfo  {journal} {Physical Review
  E}\ }\textbf {\bibinfo {volume} {80}},\ \bibinfo {pages} {010401} (\bibinfo
  {year} {2009})}\BibitemShut {NoStop}%
\bibitem [{\citenamefont {Ashkin}(1992)}]{Ashkin1}%
  \BibitemOpen
  \bibfield  {author} {\bibinfo {author} {\bibfnamefont {Arthur}\ \bibnamefont
  {Ashkin}},\ }\bibfield  {title} {\enquote {\bibinfo {title} {Forces of a
  single-beam gradient laser trap on a dielectric sphere in the ray optics
  regime},}\ }\href@noop {} {\bibfield  {journal} {\bibinfo  {journal}
  {Biophysical journal}\ }\textbf {\bibinfo {volume} {61}},\ \bibinfo {pages}
  {569--582} (\bibinfo {year} {1992})}\BibitemShut {NoStop}%
\bibitem [{\citenamefont {Sukhov}\ and\ \citenamefont
  {Dogariu}(2017)}]{Sukhov_review}%
  \BibitemOpen
  \bibfield  {author} {\bibinfo {author} {\bibfnamefont {Sergey}\ \bibnamefont
  {Sukhov}}\ and\ \bibinfo {author} {\bibfnamefont {Aristide}\ \bibnamefont
  {Dogariu}},\ }\bibfield  {title} {\enquote {\bibinfo {title}
  {Non-conservative optical forces},}\ }\href@noop {} {\bibfield  {journal}
  {\bibinfo  {journal} {Reports on Progress in Physics}\ }\textbf {\bibinfo
  {volume} {80}},\ \bibinfo {pages} {112001} (\bibinfo {year}
  {2017})}\BibitemShut {NoStop}%
\bibitem [{\citenamefont {Wu}\ \emph {et~al.}(2009)\citenamefont {Wu},
  \citenamefont {Huang}, \citenamefont {Tischer}, \citenamefont {Jonas},\ and\
  \citenamefont {Florin}}]{Florin_prl}%
  \BibitemOpen
  \bibfield  {author} {\bibinfo {author} {\bibfnamefont {Pinyu}\ \bibnamefont
  {Wu}}, \bibinfo {author} {\bibfnamefont {Rongxin}\ \bibnamefont {Huang}},
  \bibinfo {author} {\bibfnamefont {Christian}\ \bibnamefont {Tischer}},
  \bibinfo {author} {\bibfnamefont {Alexandr}\ \bibnamefont {Jonas}}, \ and\
  \bibinfo {author} {\bibfnamefont {Ernst-Ludwig}\ \bibnamefont {Florin}},\
  }\bibfield  {title} {\enquote {\bibinfo {title} {Direct measurement of the
  nonconservative force field generated by optical tweezers},}\ }\href@noop {}
  {\bibfield  {journal} {\bibinfo  {journal} {Physical review letters}\
  }\textbf {\bibinfo {volume} {103}},\ \bibinfo {pages} {108101} (\bibinfo
  {year} {2009})}\BibitemShut {NoStop}%
\bibitem [{\citenamefont {Pesce}\ \emph {et~al.}(2009)\citenamefont {Pesce},
  \citenamefont {Volpe}, \citenamefont {De~Luca}, \citenamefont {Rusciano},\
  and\ \citenamefont {Volpe}}]{Volpe_vortices_negligible}%
  \BibitemOpen
  \bibfield  {author} {\bibinfo {author} {\bibfnamefont {Giuseppe}\
  \bibnamefont {Pesce}}, \bibinfo {author} {\bibfnamefont {Giorgio}\
  \bibnamefont {Volpe}}, \bibinfo {author} {\bibfnamefont {Anna~Chiara}\
  \bibnamefont {De~Luca}}, \bibinfo {author} {\bibfnamefont {Giulia}\
  \bibnamefont {Rusciano}}, \ and\ \bibinfo {author} {\bibfnamefont {Giovanni}\
  \bibnamefont {Volpe}},\ }\bibfield  {title} {\enquote {\bibinfo {title}
  {Quantitative assessment of non-conservative radiation forces in an optical
  trap},}\ }\href@noop {} {\bibfield  {journal} {\bibinfo  {journal} {EPL
  (Europhysics Letters)}\ }\textbf {\bibinfo {volume} {86}},\ \bibinfo {pages}
  {38002} (\bibinfo {year} {2009})}\BibitemShut {NoStop}%
\bibitem [{\citenamefont {Khan}\ and\ \citenamefont
  {Sood}(2011)}]{Manas_vortexes}%
  \BibitemOpen
  \bibfield  {author} {\bibinfo {author} {\bibfnamefont {Manas}\ \bibnamefont
  {Khan}}\ and\ \bibinfo {author} {\bibfnamefont {AK}~\bibnamefont {Sood}},\
  }\bibfield  {title} {\enquote {\bibinfo {title} {Tunable brownian vortex at
  the interface},}\ }\href@noop {} {\bibfield  {journal} {\bibinfo  {journal}
  {Physical Review E}\ }\textbf {\bibinfo {volume} {83}},\ \bibinfo {pages}
  {041408} (\bibinfo {year} {2011})}\BibitemShut {NoStop}%
\bibitem [{\citenamefont {De~Cisneros}\ and\ \citenamefont
  {Hern{\'a}ndez}(2007)}]{Hernandez_engine_array}%
  \BibitemOpen
  \bibfield  {author} {\bibinfo {author} {\bibfnamefont {B~Jim{\'e}nez}\
  \bibnamefont {De~Cisneros}}\ and\ \bibinfo {author} {\bibfnamefont {A~Calvo}\
  \bibnamefont {Hern{\'a}ndez}},\ }\bibfield  {title} {\enquote {\bibinfo
  {title} {Collective working regimes for coupled heat engines},}\ }\href@noop
  {} {\bibfield  {journal} {\bibinfo  {journal} {Physical review letters}\
  }\textbf {\bibinfo {volume} {98}},\ \bibinfo {pages} {130602} (\bibinfo
  {year} {2007})}\BibitemShut {NoStop}%
\bibitem [{\citenamefont {Drescher}\ \emph {et~al.}(2010)\citenamefont
  {Drescher}, \citenamefont {Goldstein}, \citenamefont {Michel}, \citenamefont
  {Polin},\ and\ \citenamefont {Tuval}}]{Microswimmer1}%
  \BibitemOpen
  \bibfield  {author} {\bibinfo {author} {\bibfnamefont {Knut}\ \bibnamefont
  {Drescher}}, \bibinfo {author} {\bibfnamefont {Raymond~E}\ \bibnamefont
  {Goldstein}}, \bibinfo {author} {\bibfnamefont {Nicolas}\ \bibnamefont
  {Michel}}, \bibinfo {author} {\bibfnamefont {Marco}\ \bibnamefont {Polin}}, \
  and\ \bibinfo {author} {\bibfnamefont {Idan}\ \bibnamefont {Tuval}},\
  }\bibfield  {title} {\enquote {\bibinfo {title} {Direct measurement of the
  flow field around swimming microorganisms},}\ }\href@noop {} {\bibfield
  {journal} {\bibinfo  {journal} {Physical Review Letters}\ }\textbf {\bibinfo
  {volume} {105}},\ \bibinfo {pages} {168101} (\bibinfo {year}
  {2010})}\BibitemShut {NoStop}%
\bibitem [{\citenamefont {Ledesma-Aguilar}\ and\ \citenamefont
  {Yeomans}(2013)}]{Microswimmer2}%
  \BibitemOpen
  \bibfield  {author} {\bibinfo {author} {\bibfnamefont {Rodrigo}\ \bibnamefont
  {Ledesma-Aguilar}}\ and\ \bibinfo {author} {\bibfnamefont {Julia~M}\
  \bibnamefont {Yeomans}},\ }\bibfield  {title} {\enquote {\bibinfo {title}
  {Enhanced motility of a microswimmer in rigid and elastic confinement},}\
  }\href@noop {} {\bibfield  {journal} {\bibinfo  {journal} {Physical review
  letters}\ }\textbf {\bibinfo {volume} {111}},\ \bibinfo {pages} {138101}
  (\bibinfo {year} {2013})}\BibitemShut {NoStop}%
\bibitem [{\citenamefont {Pooley}\ \emph {et~al.}(2007)\citenamefont {Pooley},
  \citenamefont {Alexander},\ and\ \citenamefont {Yeomans}}]{Microswimmer4}%
  \BibitemOpen
  \bibfield  {author} {\bibinfo {author} {\bibfnamefont {CM}~\bibnamefont
  {Pooley}}, \bibinfo {author} {\bibfnamefont {GP}~\bibnamefont {Alexander}}, \
  and\ \bibinfo {author} {\bibfnamefont {JM}~\bibnamefont {Yeomans}},\
  }\bibfield  {title} {\enquote {\bibinfo {title} {Hydrodynamic interaction
  between two swimmers at low reynolds number},}\ }\href@noop {} {\bibfield
  {journal} {\bibinfo  {journal} {Physical review letters}\ }\textbf {\bibinfo
  {volume} {99}},\ \bibinfo {pages} {228103} (\bibinfo {year}
  {2007})}\BibitemShut {NoStop}%
\bibitem [{\citenamefont {F{\"a}llman}\ and\ \citenamefont
  {Axner}(1997)}]{dual_trap}%
  \BibitemOpen
  \bibfield  {author} {\bibinfo {author} {\bibfnamefont {Erik}\ \bibnamefont
  {F{\"a}llman}}\ and\ \bibinfo {author} {\bibfnamefont {Ove}\ \bibnamefont
  {Axner}},\ }\bibfield  {title} {\enquote {\bibinfo {title} {Design for fully
  steerable dual-trap optical tweezers},}\ }\href@noop {} {\bibfield  {journal}
  {\bibinfo  {journal} {Applied Optics}\ }\textbf {\bibinfo {volume} {36}},\
  \bibinfo {pages} {2107--2113} (\bibinfo {year} {1997})}\BibitemShut {NoStop}%
\bibitem [{\citenamefont {Curtis}\ \emph {et~al.}(2002)\citenamefont {Curtis},
  \citenamefont {Koss},\ and\ \citenamefont {Grier}}]{Grier_holo_tweezer1}%
  \BibitemOpen
  \bibfield  {author} {\bibinfo {author} {\bibfnamefont {Jennifer~E}\
  \bibnamefont {Curtis}}, \bibinfo {author} {\bibfnamefont {Brian~A}\
  \bibnamefont {Koss}}, \ and\ \bibinfo {author} {\bibfnamefont {David~G}\
  \bibnamefont {Grier}},\ }\bibfield  {title} {\enquote {\bibinfo {title}
  {Dynamic holographic optical tweezers},}\ }\href@noop {} {\bibfield
  {journal} {\bibinfo  {journal} {Optics communications}\ }\textbf {\bibinfo
  {volume} {207}},\ \bibinfo {pages} {169--175} (\bibinfo {year}
  {2002})}\BibitemShut {NoStop}%
\bibitem [{\citenamefont {Dufresne}\ \emph {et~al.}(2001)\citenamefont
  {Dufresne}, \citenamefont {Spalding}, \citenamefont {Dearing}, \citenamefont
  {Sheets},\ and\ \citenamefont {Grier}}]{Grier_holo_tweezer2}%
  \BibitemOpen
  \bibfield  {author} {\bibinfo {author} {\bibfnamefont {Eric~R}\ \bibnamefont
  {Dufresne}}, \bibinfo {author} {\bibfnamefont {Gabriel~C}\ \bibnamefont
  {Spalding}}, \bibinfo {author} {\bibfnamefont {Matthew~T}\ \bibnamefont
  {Dearing}}, \bibinfo {author} {\bibfnamefont {Steven~A}\ \bibnamefont
  {Sheets}}, \ and\ \bibinfo {author} {\bibfnamefont {David~G}\ \bibnamefont
  {Grier}},\ }\bibfield  {title} {\enquote {\bibinfo {title}
  {Computer-generated holographic optical tweezer arrays},}\ }\href@noop {}
  {\bibfield  {journal} {\bibinfo  {journal} {Review of Scientific
  Instruments}\ }\textbf {\bibinfo {volume} {72}},\ \bibinfo {pages}
  {1810--1816} (\bibinfo {year} {2001})}\BibitemShut {NoStop}%
\bibitem [{\citenamefont {Ashkin}\ \emph {et~al.}(1986)\citenamefont {Ashkin},
  \citenamefont {Dziedzic}, \citenamefont {Bjorkholm},\ and\ \citenamefont
  {Chu}}]{Ashkin}%
  \BibitemOpen
  \bibfield  {author} {\bibinfo {author} {\bibfnamefont {Arthur}\ \bibnamefont
  {Ashkin}}, \bibinfo {author} {\bibfnamefont {James~M}\ \bibnamefont
  {Dziedzic}}, \bibinfo {author} {\bibfnamefont {John~E}\ \bibnamefont
  {Bjorkholm}}, \ and\ \bibinfo {author} {\bibfnamefont {Steven}\ \bibnamefont
  {Chu}},\ }\bibfield  {title} {\enquote {\bibinfo {title} {Observation of a
  single-beam gradient force optical trap for dielectric particles},}\
  }\href@noop {} {\bibfield  {journal} {\bibinfo  {journal} {Optics letters}\
  }\textbf {\bibinfo {volume} {11}},\ \bibinfo {pages} {288--290} (\bibinfo
  {year} {1986})}\BibitemShut {NoStop}%
\bibitem [{\citenamefont {Kohonen}\ \emph {et~al.}(2000)\citenamefont
  {Kohonen}, \citenamefont {Karaman},\ and\ \citenamefont
  {Pashley}}]{debye_length}%
  \BibitemOpen
  \bibfield  {author} {\bibinfo {author} {\bibfnamefont {Mika~M}\ \bibnamefont
  {Kohonen}}, \bibinfo {author} {\bibfnamefont {Marilyn~E}\ \bibnamefont
  {Karaman}}, \ and\ \bibinfo {author} {\bibfnamefont {Richard~M}\ \bibnamefont
  {Pashley}},\ }\bibfield  {title} {\enquote {\bibinfo {title} {Debye length in
  multivalent electrolyte solutions},}\ }\href@noop {} {\bibfield  {journal}
  {\bibinfo  {journal} {Langmuir}\ }\textbf {\bibinfo {volume} {16}},\ \bibinfo
  {pages} {5749--5753} (\bibinfo {year} {2000})}\BibitemShut {NoStop}%
\bibitem [{\citenamefont {Meiners}\ and\ \citenamefont
  {Quake}(1999)}]{Quake_coupling}%
  \BibitemOpen
  \bibfield  {author} {\bibinfo {author} {\bibfnamefont {Jens-Christian}\
  \bibnamefont {Meiners}}\ and\ \bibinfo {author} {\bibfnamefont {Stephen~R}\
  \bibnamefont {Quake}},\ }\bibfield  {title} {\enquote {\bibinfo {title}
  {Direct measurement of hydrodynamic cross correlations between two particles
  in an external potential},}\ }\href@noop {} {\bibfield  {journal} {\bibinfo
  {journal} {Physical review letters}\ }\textbf {\bibinfo {volume} {82}},\
  \bibinfo {pages} {2211} (\bibinfo {year} {1999})}\BibitemShut {NoStop}%
\bibitem [{\citenamefont {B{\'e}rut}\ \emph
  {et~al.}(2016{\natexlab{a}})\citenamefont {B{\'e}rut}, \citenamefont
  {Imparato}, \citenamefont {Petrosyan},\ and\ \citenamefont
  {Ciliberto}}]{Ciliberto_prl}%
  \BibitemOpen
  \bibfield  {author} {\bibinfo {author} {\bibfnamefont {Antoine}\ \bibnamefont
  {B{\'e}rut}}, \bibinfo {author} {\bibfnamefont {Alberto}\ \bibnamefont
  {Imparato}}, \bibinfo {author} {\bibfnamefont {Artem}\ \bibnamefont
  {Petrosyan}}, \ and\ \bibinfo {author} {\bibfnamefont {Sergio}\ \bibnamefont
  {Ciliberto}},\ }\bibfield  {title} {\enquote {\bibinfo {title} {Stationary
  and transient fluctuation theorems for effective heat fluxes between
  hydrodynamically coupled particles in optical traps},}\ }\href@noop {}
  {\bibfield  {journal} {\bibinfo  {journal} {Physical review letters}\
  }\textbf {\bibinfo {volume} {116}},\ \bibinfo {pages} {068301} (\bibinfo
  {year} {2016}{\natexlab{a}})}\BibitemShut {NoStop}%
\bibitem [{\citenamefont {Peterman}\ \emph {et~al.}(2003)\citenamefont
  {Peterman}, \citenamefont {Gittes},\ and\ \citenamefont
  {Schmidt}}]{Schmidt_heating}%
  \BibitemOpen
  \bibfield  {author} {\bibinfo {author} {\bibfnamefont {Erwin~JG}\
  \bibnamefont {Peterman}}, \bibinfo {author} {\bibfnamefont {Frederick}\
  \bibnamefont {Gittes}}, \ and\ \bibinfo {author} {\bibfnamefont
  {Christoph~F}\ \bibnamefont {Schmidt}},\ }\bibfield  {title} {\enquote
  {\bibinfo {title} {Laser-induced heating in optical traps},}\ }\href@noop {}
  {\bibfield  {journal} {\bibinfo  {journal} {Biophysical journal}\ }\textbf
  {\bibinfo {volume} {84}},\ \bibinfo {pages} {1308--1316} (\bibinfo {year}
  {2003})}\BibitemShut {NoStop}%
\bibitem [{\citenamefont {Garrett}\ \emph {et~al.}(2018)\citenamefont
  {Garrett}, \citenamefont {Somers},\ and\ \citenamefont {Munday}}]{Casimir}%
  \BibitemOpen
  \bibfield  {author} {\bibinfo {author} {\bibfnamefont {Joseph~L}\
  \bibnamefont {Garrett}}, \bibinfo {author} {\bibfnamefont {David~AT}\
  \bibnamefont {Somers}}, \ and\ \bibinfo {author} {\bibfnamefont {Jeremy~N}\
  \bibnamefont {Munday}},\ }\bibfield  {title} {\enquote {\bibinfo {title}
  {Measurement of the casimir force between two spheres},}\ }\href@noop {}
  {\bibfield  {journal} {\bibinfo  {journal} {Physical review letters}\
  }\textbf {\bibinfo {volume} {120}},\ \bibinfo {pages} {040401} (\bibinfo
  {year} {2018})}\BibitemShut {NoStop}%
\bibitem [{\citenamefont {Korson}\ \emph {et~al.}(1969)\citenamefont {Korson},
  \citenamefont {Drost-Hansen},\ and\ \citenamefont
  {Millero}}]{water_viscosity}%
  \BibitemOpen
  \bibfield  {author} {\bibinfo {author} {\bibfnamefont {Lawrence}\
  \bibnamefont {Korson}}, \bibinfo {author} {\bibfnamefont {Walter}\
  \bibnamefont {Drost-Hansen}}, \ and\ \bibinfo {author} {\bibfnamefont
  {Frank~J}\ \bibnamefont {Millero}},\ }\bibfield  {title} {\enquote {\bibinfo
  {title} {Viscosity of water at various temperatures},}\ }\href@noop {}
  {\bibfield  {journal} {\bibinfo  {journal} {The Journal of Physical
  Chemistry}\ }\textbf {\bibinfo {volume} {73}},\ \bibinfo {pages} {34--39}
  (\bibinfo {year} {1969})}\BibitemShut {NoStop}%
\bibitem [{\citenamefont {Richerzhagen}(1996)}]{water_RI}%
  \BibitemOpen
  \bibfield  {author} {\bibinfo {author} {\bibfnamefont {Bernold}\ \bibnamefont
  {Richerzhagen}},\ }\bibfield  {title} {\enquote {\bibinfo {title}
  {Interferometer for measuring the absolute refractive index of liquid water
  as a function of temperature at 1.064 $\mu$m},}\ }\href@noop {} {\bibfield
  {journal} {\bibinfo  {journal} {Applied optics}\ }\textbf {\bibinfo {volume}
  {35}},\ \bibinfo {pages} {1650--1653} (\bibinfo {year} {1996})}\BibitemShut
  {NoStop}%
\bibitem [{\citenamefont {Sun}\ \emph {et~al.}(2010)\citenamefont {Sun},
  \citenamefont {Grier},\ and\ \citenamefont {Grosberg}}]{Grier_minimal_model}%
  \BibitemOpen
  \bibfield  {author} {\bibinfo {author} {\bibfnamefont {Bo}~\bibnamefont
  {Sun}}, \bibinfo {author} {\bibfnamefont {David~G}\ \bibnamefont {Grier}}, \
  and\ \bibinfo {author} {\bibfnamefont {Alexander~Y}\ \bibnamefont
  {Grosberg}},\ }\bibfield  {title} {\enquote {\bibinfo {title} {Minimal model
  for brownian vortexes},}\ }\href@noop {} {\bibfield  {journal} {\bibinfo
  {journal} {Physical Review E}\ }\textbf {\bibinfo {volume} {82}},\ \bibinfo
  {pages} {021123} (\bibinfo {year} {2010})}\BibitemShut {NoStop}%
\bibitem [{\citenamefont {B{\'e}rut}\ \emph
  {et~al.}(2016{\natexlab{b}})\citenamefont {B{\'e}rut}, \citenamefont
  {Imparato}, \citenamefont {Petrosyan},\ and\ \citenamefont
  {Ciliberto}}]{Ciliberto_pre}%
  \BibitemOpen
  \bibfield  {author} {\bibinfo {author} {\bibfnamefont {Antoine}\ \bibnamefont
  {B{\'e}rut}}, \bibinfo {author} {\bibfnamefont {Alberto}\ \bibnamefont
  {Imparato}}, \bibinfo {author} {\bibfnamefont {Artyom}\ \bibnamefont
  {Petrosyan}}, \ and\ \bibinfo {author} {\bibfnamefont {Sergio}\ \bibnamefont
  {Ciliberto}},\ }\bibfield  {title} {\enquote {\bibinfo {title} {Theoretical
  description of effective heat transfer between two viscously coupled
  beads},}\ }\href@noop {} {\bibfield  {journal} {\bibinfo  {journal} {Physical
  Review E}\ }\textbf {\bibinfo {volume} {94}},\ \bibinfo {pages} {052148}
  (\bibinfo {year} {2016}{\natexlab{b}})}\BibitemShut {NoStop}%
\bibitem [{\citenamefont {Herrera-Velarde}\ \emph {et~al.}(2013)\citenamefont
  {Herrera-Velarde}, \citenamefont {Eu{\'a}n-D{\'\i}az}, \citenamefont
  {C{\'o}rdoba-Vald{\'e}s},\ and\ \citenamefont
  {Castaneda-Priego}}]{herrera_jpc}%
  \BibitemOpen
  \bibfield  {author} {\bibinfo {author} {\bibfnamefont {Salvador}\
  \bibnamefont {Herrera-Velarde}}, \bibinfo {author} {\bibfnamefont {Edith~C}\
  \bibnamefont {Eu{\'a}n-D{\'\i}az}}, \bibinfo {author} {\bibfnamefont {Fidel}\
  \bibnamefont {C{\'o}rdoba-Vald{\'e}s}}, \ and\ \bibinfo {author}
  {\bibfnamefont {Ram{\'o}n}\ \bibnamefont {Castaneda-Priego}},\ }\bibfield
  {title} {\enquote {\bibinfo {title} {Hydrodynamic correlations in
  three-particle colloidal systems in harmonic traps},}\ }\href@noop {}
  {\bibfield  {journal} {\bibinfo  {journal} {Journal of Physics: Condensed
  Matter}\ }\textbf {\bibinfo {volume} {25}},\ \bibinfo {pages} {325102}
  (\bibinfo {year} {2013})}\BibitemShut {NoStop}%
\bibitem [{\citenamefont {Sekimoto}(1998)}]{Sekimoto}%
  \BibitemOpen
  \bibfield  {author} {\bibinfo {author} {\bibfnamefont {Ken}\ \bibnamefont
  {Sekimoto}},\ }\bibfield  {title} {\enquote {\bibinfo {title} {Langevin
  equation and thermodynamics},}\ }\href@noop {} {\bibfield  {journal}
  {\bibinfo  {journal} {Progress of Theoretical Physics Supplement}\ }\textbf
  {\bibinfo {volume} {130}},\ \bibinfo {pages} {17--27} (\bibinfo {year}
  {1998})}\BibitemShut {NoStop}%
\bibitem [{\citenamefont {Seifert}(2012)}]{Seifert_review}%
  \BibitemOpen
  \bibfield  {author} {\bibinfo {author} {\bibfnamefont {Udo}\ \bibnamefont
  {Seifert}},\ }\bibfield  {title} {\enquote {\bibinfo {title} {Stochastic
  thermodynamics, fluctuation theorems and molecular machines},}\ }\href@noop
  {} {\bibfield  {journal} {\bibinfo  {journal} {Reports on progress in
  physics}\ }\textbf {\bibinfo {volume} {75}},\ \bibinfo {pages} {126001}
  (\bibinfo {year} {2012})}\BibitemShut {NoStop}%
\bibitem [{\citenamefont {Bechinger}\ \emph {et~al.}(2016)\citenamefont
  {Bechinger}, \citenamefont {Di~Leonardo}, \citenamefont {L{\"o}wen},
  \citenamefont {Reichhardt}, \citenamefont {Volpe},\ and\ \citenamefont
  {Volpe}}]{Bechinger_rmp}%
  \BibitemOpen
  \bibfield  {author} {\bibinfo {author} {\bibfnamefont {Clemens}\ \bibnamefont
  {Bechinger}}, \bibinfo {author} {\bibfnamefont {Roberto}\ \bibnamefont
  {Di~Leonardo}}, \bibinfo {author} {\bibfnamefont {Hartmut}\ \bibnamefont
  {L{\"o}wen}}, \bibinfo {author} {\bibfnamefont {Charles}\ \bibnamefont
  {Reichhardt}}, \bibinfo {author} {\bibfnamefont {Giorgio}\ \bibnamefont
  {Volpe}}, \ and\ \bibinfo {author} {\bibfnamefont {Giovanni}\ \bibnamefont
  {Volpe}},\ }\bibfield  {title} {\enquote {\bibinfo {title} {Active particles
  in complex and crowded environments},}\ }\href@noop {} {\bibfield  {journal}
  {\bibinfo  {journal} {Reviews of Modern Physics}\ }\textbf {\bibinfo {volume}
  {88}},\ \bibinfo {pages} {045006} (\bibinfo {year} {2016})}\BibitemShut
  {NoStop}%
\bibitem [{\citenamefont {Palacci}\ \emph {et~al.}(2010)\citenamefont
  {Palacci}, \citenamefont {Cottin-Bizonne}, \citenamefont {Ybert},\ and\
  \citenamefont {Bocquet}}]{Bocquet}%
  \BibitemOpen
  \bibfield  {author} {\bibinfo {author} {\bibfnamefont {J{\'e}r{\'e}mie}\
  \bibnamefont {Palacci}}, \bibinfo {author} {\bibfnamefont {C{\'e}cile}\
  \bibnamefont {Cottin-Bizonne}}, \bibinfo {author} {\bibfnamefont
  {Christophe}\ \bibnamefont {Ybert}}, \ and\ \bibinfo {author} {\bibfnamefont
  {Lyd{\'e}ric}\ \bibnamefont {Bocquet}},\ }\bibfield  {title} {\enquote
  {\bibinfo {title} {Sedimentation and effective temperature of active
  colloidal suspensions},}\ }\href@noop {} {\bibfield  {journal} {\bibinfo
  {journal} {Physical Review Letters}\ }\textbf {\bibinfo {volume} {105}},\
  \bibinfo {pages} {088304} (\bibinfo {year} {2010})}\BibitemShut {NoStop}%
\bibitem [{\citenamefont {Bricard}\ \emph {et~al.}(2013)\citenamefont
  {Bricard}, \citenamefont {Caussin}, \citenamefont {Desreumaux}, \citenamefont
  {Dauchot},\ and\ \citenamefont {Bartolo}}]{Bartolo}%
  \BibitemOpen
  \bibfield  {author} {\bibinfo {author} {\bibfnamefont {Antoine}\ \bibnamefont
  {Bricard}}, \bibinfo {author} {\bibfnamefont {Jean-Baptiste}\ \bibnamefont
  {Caussin}}, \bibinfo {author} {\bibfnamefont {Nicolas}\ \bibnamefont
  {Desreumaux}}, \bibinfo {author} {\bibfnamefont {Olivier}\ \bibnamefont
  {Dauchot}}, \ and\ \bibinfo {author} {\bibfnamefont {Denis}\ \bibnamefont
  {Bartolo}},\ }\bibfield  {title} {\enquote {\bibinfo {title} {Emergence of
  macroscopic directed motion in populations of motile colloids},}\ }\href@noop
  {} {\bibfield  {journal} {\bibinfo  {journal} {Nature}\ }\textbf {\bibinfo
  {volume} {503}},\ \bibinfo {pages} {95--98} (\bibinfo {year}
  {2013})}\BibitemShut {NoStop}%
\bibitem [{\citenamefont {Buzhardt}\ and\ \citenamefont
  {Tallapragada}(2019)}]{Magnetic_rotors}%
  \BibitemOpen
  \bibfield  {author} {\bibinfo {author} {\bibfnamefont {Jake}\ \bibnamefont
  {Buzhardt}}\ and\ \bibinfo {author} {\bibfnamefont {Phanindra}\ \bibnamefont
  {Tallapragada}},\ }\bibfield  {title} {\enquote {\bibinfo {title} {Dynamics
  of groups of magnetically driven artificial microswimmers},}\ }\href@noop {}
  {\bibfield  {journal} {\bibinfo  {journal} {Physical Review E}\ }\textbf
  {\bibinfo {volume} {100}},\ \bibinfo {pages} {033106} (\bibinfo {year}
  {2019})}\BibitemShut {NoStop}%
\bibitem [{\citenamefont {Bayati}\ and\ \citenamefont
  {Najafi}(2016)}]{Microswimmer3}%
  \BibitemOpen
  \bibfield  {author} {\bibinfo {author} {\bibfnamefont {Parvin}\ \bibnamefont
  {Bayati}}\ and\ \bibinfo {author} {\bibfnamefont {Ali}\ \bibnamefont
  {Najafi}},\ }\bibfield  {title} {\enquote {\bibinfo {title} {Dynamics of two
  interacting active janus particles},}\ }\href@noop {} {\bibfield  {journal}
  {\bibinfo  {journal} {The Journal of chemical physics}\ }\textbf {\bibinfo
  {volume} {144}},\ \bibinfo {pages} {134901} (\bibinfo {year}
  {2016})}\BibitemShut {NoStop}%
\bibitem [{\citenamefont {Solovev}\ \emph {et~al.}(2013)\citenamefont
  {Solovev}, \citenamefont {Sanchez},\ and\ \citenamefont {Schmidt}}]{Sanchez}%
  \BibitemOpen
  \bibfield  {author} {\bibinfo {author} {\bibfnamefont {Alexander~A}\
  \bibnamefont {Solovev}}, \bibinfo {author} {\bibfnamefont {Samuel}\
  \bibnamefont {Sanchez}}, \ and\ \bibinfo {author} {\bibfnamefont {Oliver~G}\
  \bibnamefont {Schmidt}},\ }\bibfield  {title} {\enquote {\bibinfo {title}
  {Collective behaviour of self-propelled catalytic micromotors},}\ }\href@noop
  {} {\bibfield  {journal} {\bibinfo  {journal} {Nanoscale}\ }\textbf {\bibinfo
  {volume} {5}},\ \bibinfo {pages} {1284--1293} (\bibinfo {year}
  {2013})}\BibitemShut {NoStop}%
\bibitem [{\citenamefont {Driscoll}\ and\ \citenamefont
  {Delmotte}(2019)}]{Hydrodynamics2}%
  \BibitemOpen
  \bibfield  {author} {\bibinfo {author} {\bibfnamefont {Michelle}\
  \bibnamefont {Driscoll}}\ and\ \bibinfo {author} {\bibfnamefont {Blaise}\
  \bibnamefont {Delmotte}},\ }\bibfield  {title} {\enquote {\bibinfo {title}
  {Leveraging collective effects in externally driven colloidal suspensions:
  Experiments and simulations},}\ }\href@noop {} {\bibfield  {journal}
  {\bibinfo  {journal} {Current opinion in colloid \& interface science}\
  }\textbf {\bibinfo {volume} {40}},\ \bibinfo {pages} {42--57} (\bibinfo
  {year} {2019})}\BibitemShut {NoStop}%
\end{thebibliography}%


\begin{thebibliography}{12}%
\makeatletter
\providecommand \@ifxundefined [1]{%
 \@ifx{#1\undefined}
}%
\providecommand \@ifnum [1]{%
 \ifnum #1\expandafter \@firstoftwo
 \else \expandafter \@secondoftwo
 \fi
}%
\providecommand \@ifx [1]{%
 \ifx #1\expandafter \@firstoftwo
 \else \expandafter \@secondoftwo
 \fi
}%
\providecommand \natexlab [1]{#1}%
\providecommand \enquote  [1]{``#1''}%
\providecommand \bibnamefont  [1]{#1}%
\providecommand \bibfnamefont [1]{#1}%
\providecommand \citenamefont [1]{#1}%
\providecommand \href@noop [0]{\@secondoftwo}%
\providecommand \href [0]{\begingroup \@sanitize@url \@href}%
\providecommand \@href[1]{\@@startlink{#1}\@@href}%
\providecommand \@@href[1]{\endgroup#1\@@endlink}%
\providecommand \@sanitize@url [0]{\catcode `\\12\catcode `\$12\catcode
  `\&12\catcode `\#12\catcode `\^12\catcode `\_12\catcode `\%12\relax}%
\providecommand \@@startlink[1]{}%
\providecommand \@@endlink[0]{}%
\providecommand \url  [0]{\begingroup\@sanitize@url \@url }%
\providecommand \@url [1]{\endgroup\@href {#1}{\urlprefix }}%
\providecommand \urlprefix  [0]{URL }%
\providecommand \Eprint [0]{\href }%
\providecommand \doibase [0]{http://dx.doi.org/}%
\providecommand \selectlanguage [0]{\@gobble}%
\providecommand \bibinfo  [0]{\@secondoftwo}%
\providecommand \bibfield  [0]{\@secondoftwo}%
\providecommand \translation [1]{[#1]}%
\providecommand \BibitemOpen [0]{}%
\providecommand \bibitemStop [0]{}%
\providecommand \bibitemNoStop [0]{.\EOS\space}%
\providecommand \EOS [0]{\spacefactor3000\relax}%
\providecommand \BibitemShut  [1]{\csname bibitem#1\endcsname}%
\let\auto@bib@innerbib\@empty
\bibitem [{\citenamefont {Crocker}(1997)}]{Crocker_coupling}%
  \BibitemOpen
  \bibfield  {author} {\bibinfo {author} {\bibfnamefont {J.~C.}\ \bibnamefont
  {Crocker}},\ }\href@noop {} {\bibfield  {journal} {\bibinfo  {journal} {The
  Journal of chemical physics}\ }\textbf {\bibinfo {volume} {106}},\ \bibinfo
  {pages} {2837} (\bibinfo {year} {1997})}\BibitemShut {NoStop}%
\bibitem [{\citenamefont {Pesce}\ \emph {et~al.}(2014)\citenamefont {Pesce},
  \citenamefont {Volpe}, \citenamefont {Volpe},\ and\ \citenamefont
  {Sasso}}]{Volpe_coupling}%
  \BibitemOpen
  \bibfield  {author} {\bibinfo {author} {\bibfnamefont {G.}~\bibnamefont
  {Pesce}}, \bibinfo {author} {\bibfnamefont {G.}~\bibnamefont {Volpe}},
  \bibinfo {author} {\bibfnamefont {G.}~\bibnamefont {Volpe}}, \ and\ \bibinfo
  {author} {\bibfnamefont {A.}~\bibnamefont {Sasso}},\ }\href@noop {}
  {\bibfield  {journal} {\bibinfo  {journal} {Physical Review E}\ }\textbf
  {\bibinfo {volume} {90}},\ \bibinfo {pages} {042309} (\bibinfo {year}
  {2014})}\BibitemShut {NoStop}%
\bibitem [{\citenamefont {Meiners}\ and\ \citenamefont
  {Quake}(1999)}]{Quake_coupling}%
  \BibitemOpen
  \bibfield  {author} {\bibinfo {author} {\bibfnamefont {J.-C.}\ \bibnamefont
  {Meiners}}\ and\ \bibinfo {author} {\bibfnamefont {S.~R.}\ \bibnamefont
  {Quake}},\ }\href@noop {} {\bibfield  {journal} {\bibinfo  {journal}
  {Physical review letters}\ }\textbf {\bibinfo {volume} {82}},\ \bibinfo
  {pages} {2211} (\bibinfo {year} {1999})}\BibitemShut {NoStop}%
\bibitem [{\citenamefont {de~Messieres}\ \emph {et~al.}(2011)\citenamefont
  {de~Messieres}, \citenamefont {Denesyuk},\ and\ \citenamefont
  {La~Porta}}]{Vortex_noise}%
  \BibitemOpen
  \bibfield  {author} {\bibinfo {author} {\bibfnamefont {M.}~\bibnamefont
  {de~Messieres}}, \bibinfo {author} {\bibfnamefont {N.~A.}\ \bibnamefont
  {Denesyuk}}, \ and\ \bibinfo {author} {\bibfnamefont {A.}~\bibnamefont
  {La~Porta}},\ }\href@noop {} {\bibfield  {journal} {\bibinfo  {journal}
  {Physical Review E}\ }\textbf {\bibinfo {volume} {84}},\ \bibinfo {pages}
  {031108} (\bibinfo {year} {2011})}\BibitemShut {NoStop}%
\bibitem [{\citenamefont {Blickle}\ and\ \citenamefont
  {Bechinger}(2012)}]{Bechinger_Nature}%
  \BibitemOpen
  \bibfield  {author} {\bibinfo {author} {\bibfnamefont {V.}~\bibnamefont
  {Blickle}}\ and\ \bibinfo {author} {\bibfnamefont {C.}~\bibnamefont
  {Bechinger}},\ }\href@noop {} {\bibfield  {journal} {\bibinfo  {journal}
  {Nature Physics}\ }\textbf {\bibinfo {volume} {8}},\ \bibinfo {pages} {143}
  (\bibinfo {year} {2012})}\BibitemShut {NoStop}%
\bibitem [{\citenamefont {Krishnamurthy}\ \emph {et~al.}(2016)\citenamefont
  {Krishnamurthy}, \citenamefont {Ghosh}, \citenamefont {Chatterji},
  \citenamefont {Ganapathy},\ and\ \citenamefont {Sood}}]{active_engine}%
  \BibitemOpen
  \bibfield  {author} {\bibinfo {author} {\bibfnamefont {S.}~\bibnamefont
  {Krishnamurthy}}, \bibinfo {author} {\bibfnamefont {S.}~\bibnamefont
  {Ghosh}}, \bibinfo {author} {\bibfnamefont {D.}~\bibnamefont {Chatterji}},
  \bibinfo {author} {\bibfnamefont {R.}~\bibnamefont {Ganapathy}}, \ and\
  \bibinfo {author} {\bibfnamefont {A.}~\bibnamefont {Sood}},\ }\href@noop {}
  {\bibfield  {journal} {\bibinfo  {journal} {Nature Physics}\ }\textbf
  {\bibinfo {volume} {12}},\ \bibinfo {pages} {1134} (\bibinfo {year}
  {2016})}\BibitemShut {NoStop}%
\bibitem [{\citenamefont {B{\'e}rut}\ \emph
  {et~al.}(2016{\natexlab{a}})\citenamefont {B{\'e}rut}, \citenamefont
  {Imparato}, \citenamefont {Petrosyan},\ and\ \citenamefont
  {Ciliberto}}]{Ciliberto_prl}%
  \BibitemOpen
  \bibfield  {author} {\bibinfo {author} {\bibfnamefont {A.}~\bibnamefont
  {B{\'e}rut}}, \bibinfo {author} {\bibfnamefont {A.}~\bibnamefont {Imparato}},
  \bibinfo {author} {\bibfnamefont {A.}~\bibnamefont {Petrosyan}}, \ and\
  \bibinfo {author} {\bibfnamefont {S.}~\bibnamefont {Ciliberto}},\ }\href@noop
  {} {\bibfield  {journal} {\bibinfo  {journal} {Physical review letters}\
  }\textbf {\bibinfo {volume} {116}},\ \bibinfo {pages} {068301} (\bibinfo
  {year} {2016}{\natexlab{a}})}\BibitemShut {NoStop}%
\bibitem [{\citenamefont {Pesce}\ \emph {et~al.}(2009)\citenamefont {Pesce},
  \citenamefont {Volpe}, \citenamefont {De~Luca}, \citenamefont {Rusciano},\
  and\ \citenamefont {Volpe}}]{Volpe_vortices_negligible}%
  \BibitemOpen
  \bibfield  {author} {\bibinfo {author} {\bibfnamefont {G.}~\bibnamefont
  {Pesce}}, \bibinfo {author} {\bibfnamefont {G.}~\bibnamefont {Volpe}},
  \bibinfo {author} {\bibfnamefont {A.~C.}\ \bibnamefont {De~Luca}}, \bibinfo
  {author} {\bibfnamefont {G.}~\bibnamefont {Rusciano}}, \ and\ \bibinfo
  {author} {\bibfnamefont {G.}~\bibnamefont {Volpe}},\ }\href@noop {}
  {\bibfield  {journal} {\bibinfo  {journal} {EPL (Europhysics Letters)}\
  }\textbf {\bibinfo {volume} {86}},\ \bibinfo {pages} {38002} (\bibinfo {year}
  {2009})}\BibitemShut {NoStop}%
\bibitem [{\citenamefont {B{\'e}rut}\ \emph
  {et~al.}(2016{\natexlab{b}})\citenamefont {B{\'e}rut}, \citenamefont
  {Imparato}, \citenamefont {Petrosyan},\ and\ \citenamefont
  {Ciliberto}}]{Ciliberto_pre}%
  \BibitemOpen
  \bibfield  {author} {\bibinfo {author} {\bibfnamefont {A.}~\bibnamefont
  {B{\'e}rut}}, \bibinfo {author} {\bibfnamefont {A.}~\bibnamefont {Imparato}},
  \bibinfo {author} {\bibfnamefont {A.}~\bibnamefont {Petrosyan}}, \ and\
  \bibinfo {author} {\bibfnamefont {S.}~\bibnamefont {Ciliberto}},\ }\href@noop
  {} {\bibfield  {journal} {\bibinfo  {journal} {Physical Review E}\ }\textbf
  {\bibinfo {volume} {94}},\ \bibinfo {pages} {052148} (\bibinfo {year}
  {2016}{\natexlab{b}})}\BibitemShut {NoStop}%
\bibitem [{\citenamefont {Herrera-Velarde}\ \emph {et~al.}(2013)\citenamefont
  {Herrera-Velarde}, \citenamefont {Eu{\'a}n-D{\'\i}az}, \citenamefont
  {C{\'o}rdoba-Vald{\'e}s},\ and\ \citenamefont
  {Castaneda-Priego}}]{herrera_jpc}%
  \BibitemOpen
  \bibfield  {author} {\bibinfo {author} {\bibfnamefont {S.}~\bibnamefont
  {Herrera-Velarde}}, \bibinfo {author} {\bibfnamefont {E.~C.}\ \bibnamefont
  {Eu{\'a}n-D{\'\i}az}}, \bibinfo {author} {\bibfnamefont {F.}~\bibnamefont
  {C{\'o}rdoba-Vald{\'e}s}}, \ and\ \bibinfo {author} {\bibfnamefont
  {R.}~\bibnamefont {Castaneda-Priego}},\ }\href@noop {} {\bibfield  {journal}
  {\bibinfo  {journal} {Journal of Physics: Condensed Matter}\ }\textbf
  {\bibinfo {volume} {25}},\ \bibinfo {pages} {325102} (\bibinfo {year}
  {2013})}\BibitemShut {NoStop}%
\bibitem [{\citenamefont {Bra{\'n}ka}\ and\ \citenamefont
  {Heyes}(1998)}]{branka_algorithms}%
  \BibitemOpen
  \bibfield  {author} {\bibinfo {author} {\bibfnamefont {A.}~\bibnamefont
  {Bra{\'n}ka}}\ and\ \bibinfo {author} {\bibfnamefont {D.}~\bibnamefont
  {Heyes}},\ }\href@noop {} {\bibfield  {journal} {\bibinfo  {journal}
  {Physical Review E}\ }\textbf {\bibinfo {volume} {58}},\ \bibinfo {pages}
  {2611} (\bibinfo {year} {1998})}\BibitemShut {NoStop}%
\bibitem [{\citenamefont {Bra{\'n}ka}\ and\ \citenamefont
  {Heyes}(1999)}]{branka_algorithms1}%
  \BibitemOpen
  \bibfield  {author} {\bibinfo {author} {\bibfnamefont {A.}~\bibnamefont
  {Bra{\'n}ka}}\ and\ \bibinfo {author} {\bibfnamefont {D.~M.}\ \bibnamefont
  {Heyes}},\ }\href@noop {} {\bibfield  {journal} {\bibinfo  {journal}
  {Physical Review E}\ }\textbf {\bibinfo {volume} {60}},\ \bibinfo {pages}
  {2381} (\bibinfo {year} {1999})}\BibitemShut {NoStop}%
\end{thebibliography}%

\newpage

\newpage

\newpage

\newpage

\newpage

\end{document}


\title{Supplementary information on Synergistic action in colloidal heat engines coupled by non-conservative flows}
\author{Sudeesh Krishnamurthy}
\affiliation{Department of Physics, Indian Institute of Science, Bangalore - 560012, INDIA}
\author{Rajesh Ganapathy}
\affiliation{International Centre for Materials Science, Jawaharlal Nehru Centre for Advanced Scientific Research, Jakkur, Bangalore - 560064, INDIA}
\affiliation{Sheikh Saqr Laboratory, Jawaharlal Nehru Centre for Advanced Scientific Research, Jakkur, Bangalore - 560064, INDIA}
\author{A. K. Sood}
\affiliation{Department of Physics, Indian Institute of Science, Bangalore - 560012, INDIA}
\affiliation{International Centre for Materials Science, Jawaharlal Nehru Centre for Advanced Scientific Research, Jakkur, Bangalore - 560064, INDIA}
\date{\today}

\maketitle
\newpage

\section{Correlations and vortexes}
\begin{figure}[!t]
\centering
\includegraphics[width=0.99\textwidth]{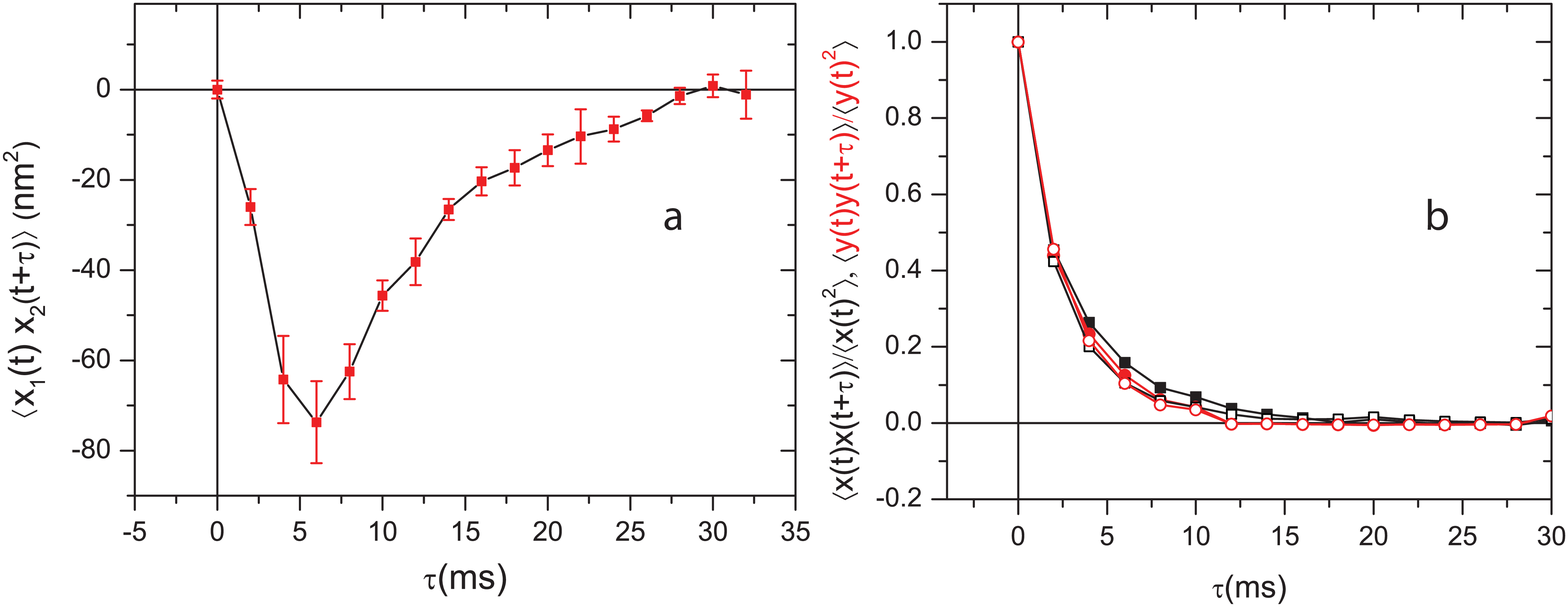}
\caption{\textbf{Cross-correlation of particle displacements. a} Cross-correlation, $\langle x_1(t) x_2(t+\tau)\rangle$ between displacements $x1,x2$ of particles trapped in T1 and T2 respectively show strong anticorrelation at $\approx 6ms$. The traps are separated by a distance of $3.6\mu m$ and the solution temperature was maintained at 300K.The correlations, however last up to 25ms. \textbf{b} shows auto-correlations along both x and y axes $\frac{\langle x_i(t)x_i(t+\tau)\rangle}{\langle x_i(t)^2\rangle}$ (black square symbols) and  $\frac{\langle y_i(t)y_i(t+\tau)\rangle}{\langle y_i(t)^2\rangle}$ (red circular symbols) for particles in the trap L1 (closed symbols) and L2 (open symbols).}
\label {FigS1}
\end{figure}

In the low Reynolds number regime, as in our experiment, velocity correlations of a colloidal particle in an isolated optical trap are known to decay sharply in timescales of the order of $\mu s$. Although motion of the colloidal microsphere through the fluid does generate hydrodynamic flows \cite{Crocker_coupling, Volpe_coupling}, accounting for the changes in particle motion due to the feedback from these flows is necessary only at sufficiently low viscosity and is negligible in our experiments. Nevertheless, in the presence of another particle in close proximity, hydrodynamic flows could mutually affect each other and with interactions dictated by the Oseen tensor,  it can be shown that these could result in an anti-correlated motion \cite{Quake_coupling} along the line joining them. However, as constrained by the fluctuation dissipation relation, these flows do not lead to any net energy transfer, but, the noise correlations do persist longer. We plot the position cross-correlation, $\langle x_1(t) x_2(t+\tau)\rangle$, where $x_1,x_2$ are the respective displacements of the particles in T1 and T2 along the line joining them, the x-axis for our experiment in Fig. S\ref{FigS1} a. As predicted in earlier experiments \cite{Quake_coupling}, we observe an initial decrease in $\langle x_1(t) x_2(t+\tau)\rangle$, with maximum anti-correlation at $\approx 6ms$. Also, $\langle x_1(t) x_2(t+\tau)\rangle$ is significantly negative even at large $\tau \approx 25ms$. We exploit this to inject a notable reflux of energy from the ensuing vortexes \cite{Vortex_noise}, the active driving in our system to the particle motion. Also, this energy transfer can now be manipulated using solution viscosity as a switch to create reservoirs of high and low energy addition mimicking temperature/activity. In our experiments, we emulated the standard protocol of microengines \cite{Bechinger_Nature,active_engine} to operate a Stirling cycle simultaneously on both traps using this inter-trap heat transfer to generate work.

Also, in the main paper, we discussed that the position of the particle in the x-y plane decides the amount of light scattered, the primary source of non-equilibrium driving in our system. We argued that since the positions in the x-y plane are approximately exponentially correlated, we used a similar profile for noise correlations in calculating thermodynamic quantities. In Fig. S\ref{FigS1} b, we plot spatial auto-correlations along both x and y axes for both the trapped particles. Due to the symmetry of the optical trap, its stiffness along x and y axes are almost equal and the position correlations along them are similar. Further, even though the laser intensities $S(L2) = 1.33S(L1)$, difference in their stiffness $\approx 0.8 pN/\mu m$ did not produce significantly different correlations. While analytically, the spatial auto-correlations are known to decay as double exponential \cite{Ciliberto_prl}, we observed that this could be approximated for simplifying our calculations to a single exponential with a time constant $\approx 5-6 ms$. The stochastic thermodynamics model used in our experiments to calculate thermodynamic quantities was based on this assumption. The Brownian dynamics simulations in Fig. 3f of the main paper used an exponentially correlated noise that decayed over the same order $= 10ms$ of timescales.

\section{Changes in the vortex behavior between the hot and cold reservoirs}
\begin{figure}[!t]
\centering
\includegraphics[width=0.99\textwidth]{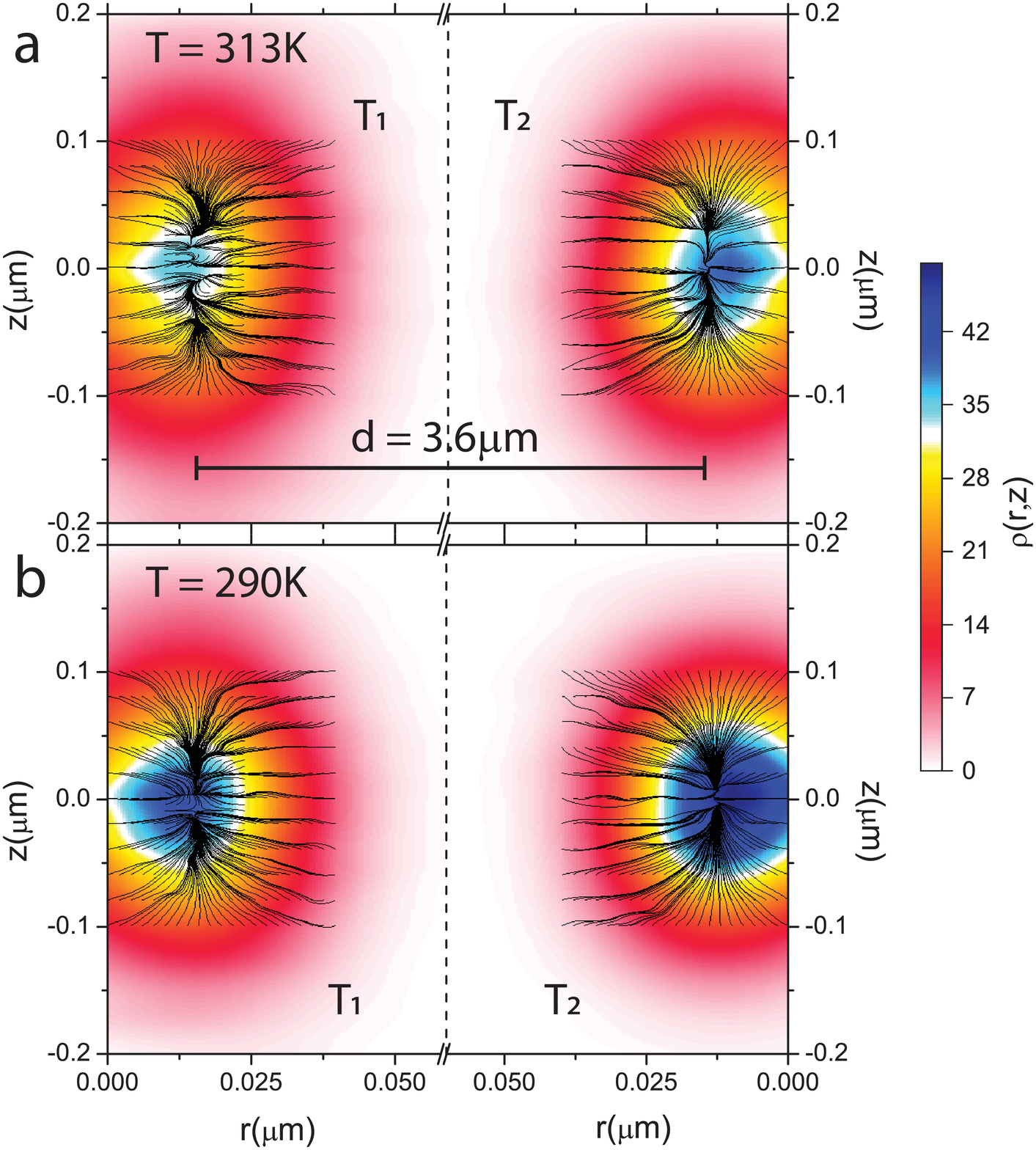}
\caption{\textbf{Brownian vortexes in isolated traps} \textbf{a} and \textbf{b} show streamlines of average particle particle motion plotted in the r-z plane for both T1 and T2 before another identical particle was introduced in the adjacent trap in hot and cold reservoirs respectively. The total laser power, $S$, was maintained at the maximum of $\approx 540 mW$ with $S(T2)/S(T1) = 1.33$ during the experiment. The background image represents the probability density of the particle, $\rho(r,z)$ color coded as shown in the color bar.}
\label {FigS2}
\end{figure}

In the main manuscript we described that the vortexes are enhanced due to mutual interaction when another particle is introduced in the adjacent trap. Nevertheless, the increase in $E_t$ due to this depends on the ambient parameters such as viscosity, separation between the particles etc. as described in Fig.s 1 c and d. Here, to reaffirm our results, we demonstrate that the same is also true of the vortexes as well. To this extent, we observe the streamlines of particle displacements as in Fig. 2 for $P = 540mW$ where, maximum difference in $E_t$ between $T = 313K$ and  $T = 290K$ was observed in our experiments. Fig. S\ref{FigS2} describes the streamlines before the addition of another particle into the adjacent trap at both the temperatures $T = 313K$ and $T = 290K$. As in Fig. 2, we observe that the particles released at the extremities converged to a region close to $z=0$ plane but at a finite r. The ensuing vortexes in Fig. S\ref{FigS2}, however, are more pronounced in the trajectories at $T = 313K$ as opposed to $T = 290K$. Thus, as anticipated by our observations, the presence of prominent vortexes in isolated traps leads to a greater $E_t$. The change in the vortex pattern in turn is due to the difference in the viscosity of the suspending medium.

\section{Probability distributions along x and y-directions}
\begin{figure}[tbp]
\centering
\includegraphics[width=0.8\textwidth]{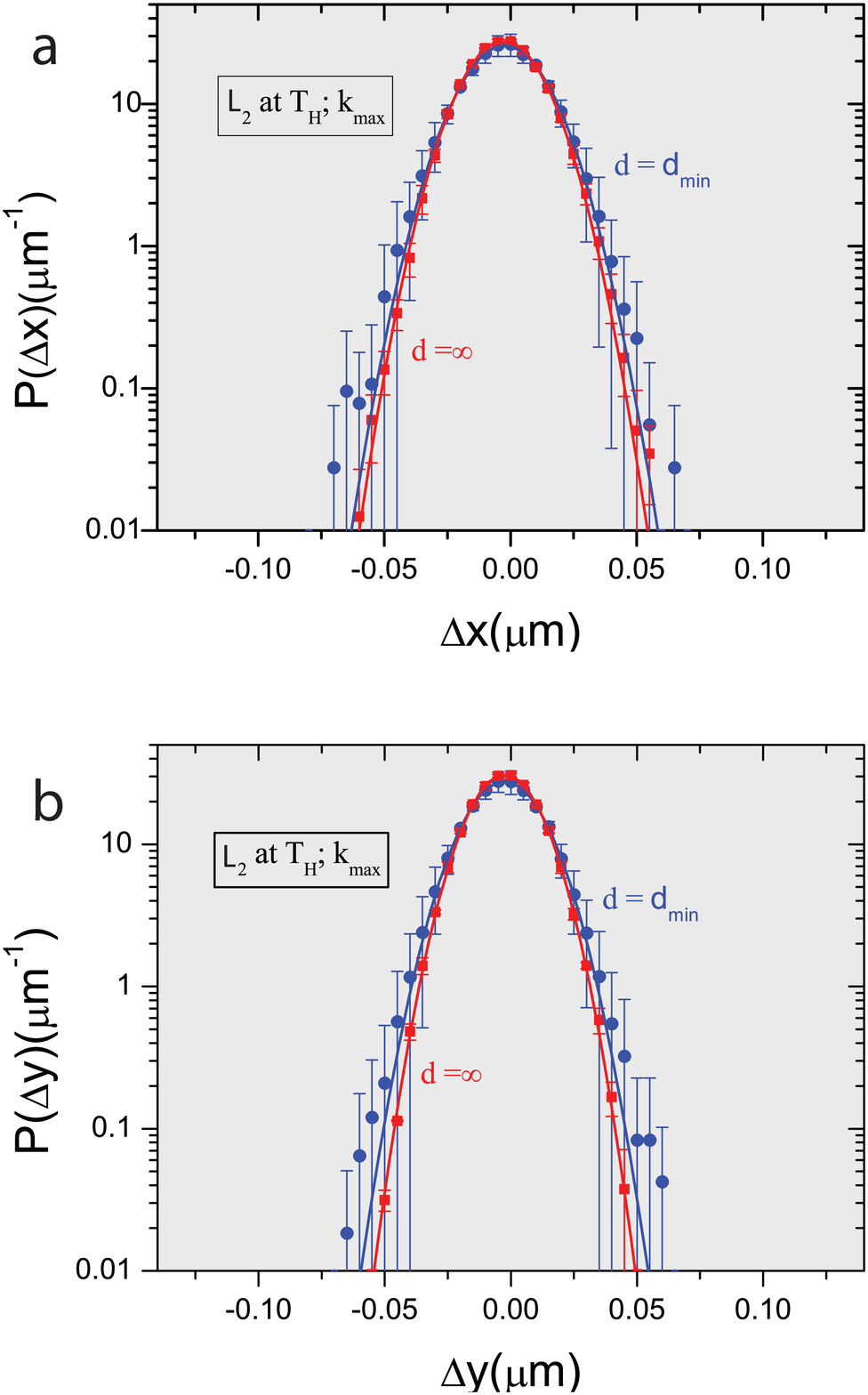}
\caption{\textbf{Probability distributions along x and y directions} \textbf{a} and \textbf{b} show $P(\Delta x)$ and $P(\Delta y)$ respectively for a particle in T2 before(red squares) and after(blue circles) introducing an identical particle in T1 at the same extreme conditions in Fig. 5 of the main paper. The solid lines represent Gaussian fits for the same.}
\label {FigS3}
\end{figure}

In the main manuscript, we considered the motion only along the z-direction to calculate $T_{act}$, since the addition of another particle in the adjacent trap causes negligible changes in $P(\Delta x)$ and $P(\Delta y)$ as seen in Fig. 1 b of the main paper. To reaffirm that they are indeed insignificant, we plot $P(\Delta x)$ and $P(\Delta y)$ for T2 before and after introducing the other colloidal microsphere in the neighboring trap in Fig. S\ref{FigS3} at the same extreme conditions in Fig. 5 of the main paper. From the Fig. S\ref{FigS3}, the small changes in $P(\Delta x)$ and $P(\Delta y)$ are within the limits of experimental error. Although there is a slight change of statistics in the tails, the combined increase in energy enclosed by $P(\Delta x)$ and $P(\Delta y)$, on coupling, was found to be $\approx 0.05 k_BT_C$. Thus, most of the contributions for an elevated $T_{act}$ arise from the changes in the z-direction.

\section{Stochastic thermodynamics of coupled engines}
To calculate work and efficiency of our engines, we used the framework of stochastic thermodynamics. The equation of motion for an isolated sphere in an optical trap is 
\begin{equation}
\gamma (\dot{\textbf{r}}-\acute{v_1}(\textbf{r},S)) = -\nabla U_1 + F_{1s}(\textbf{r},S) +\xi_1(t)
\end{equation}
where, $\gamma = 6\pi\mu(\sigma/2)$ is the friction co-efficient, U is the conservative potential corresponding to the gradient force, $F_{1s}$ is the non-conservative scattering force, $\xi_1$ is the thermal noise and $\acute{v_1}(\textbf{r},S)$ is the fluid flow generated by the dissipation of energy input by $F_{1s}$. In our experiments, we fix the co-ordinate system such that the laser propagates along the z-axis. Apart from displacing the mean position away from the origin, as observed in Fig. 3 of the main paper, $F_{1s}$ and $\acute{v_1}(\textbf{r},S)$ are negligible along the x and y axes in comparison with the gradient force. The equation of motion along the z-axis is
\begin{equation}
\gamma \dot{z} = -k_{1z} z + [F_{1s}(\textbf{r},S) +\gamma\acute{v_1}(\textbf{r},S)] +\xi(t)
\end{equation}
As noted in the main paper, the feedback of energy from the terms in the parenthesis to the particle motion (typically up to $4\%$ of the thermal energy) can be neglected in most experiments \cite{Volpe_vortices_negligible} and can be treated as perturbations over the equilibrium behavior. Adding another identical particle to an adjacent optical trap induces hydrodynamic interactions between the two particles. The resulting equations of motion along the z-axis would be
\begin{equation}
\begin{split}
\gamma \dot{z_1} = -k_{1z} z_1 +\epsilon (-k_{2z} z_2+F_{2s})+ [F_{1s}(\mathbf{r_1},S_1) +\gamma\acute{v_1}] +\xi_1(t)\\
\gamma \dot{z_2} = -k_{2z} z_2 +\epsilon (-k_{1z} z_1+F_{1s})+ [F_{2s}(\mathbf{r_2},S_2) +\gamma\acute{v_2}] +\xi_2(t)
\end{split}
\end{equation}
where $\epsilon$ is the hydrodynamic coupling constant and can be calculated from the Rotne-Prager diffusion tensor to be $\epsilon = \frac{3\sigma}{4d} + \frac{\sigma^3}{2d^3}$ (For similar treatments of coupled hydrodynamics see \cite{Ciliberto_prl, Ciliberto_pre, herrera_jpc}). Proceeding further from this would require an explicit form for $F_{is}$ and $\acute{v_i}$. While studies in the past have tried to arrive at closed form expressions for these, apart from addressing general features such as reversal of circulatory flows etc., such approximations have been unable to capture the particle dynamics. In difference with these studies, motivated by the chaotic nature of the trajectories that we observed in Fig. 2 of the main paper, we approximate these terms to a non-equilibrium noise, $\alpha_i(t) = F_{is} + \epsilon F_{js} + \gamma\acute{v_i}$ with $i\neq j$. The equations of motion would then be
\begin{equation}
\begin{split}
\gamma \dot{z_1} = -k_{1z} z_1 -\epsilon k_{2z} z_2 + \alpha_1(t) +\xi_1(t)\\
\gamma \dot{z_2} = -k_{2z} z_2 -\epsilon k_{1z} z_1 + \alpha_2(t) +\xi_2(t)
\end{split}
\label{eq_motion1}
\end{equation}
Since the observed mean displacement along the z-axis was negligible, we impose $\langle\alpha_i(t)\rangle = 0$. However, since previous studies have noted that the position-auto correlation function along x and y directions decayed as a double exponential \cite{Ciliberto_prl, Ciliberto_pre} (Plotted in Fig. S\ref{FigS1}b for our data), and the scattering force crucially depends on the particle position, as a first order approximation, we choose $\langle\alpha_i(t)\alpha_i(t')\rangle = A_i e^{-\frac{t-t'}{t_{\alpha}}}$. Analogous to the case of the isolated trapped particle, we define the heat transferred to the particle i, $Q_i$ in a time interval $\tau$ as
\begin{equation}
Q_i = -\int_{0}^{\tau}{(-\gamma \dot{z_i}+\alpha_i(t) +\xi_i(t))\dot{z_i}dt} 
\end{equation}
Such a definition treats $\alpha_i$ as a driving active noise analogous to an active Brownian particle and any energy received from $F_{is}$ and $\acute{v_i}(\textbf{r},S)$ is similar to that received from the thermal reservoir except through different statistics. Using \ref{eq_motion1}, we can rewrite this as  
\begin{equation}
Q_i = -\int_{0}^{\tau}{(k_{iz} z_i +\epsilon k_{jz} z_j)\dot{z_i}dt}; i\neq j
\end{equation}
Following Berut et.al. \cite{Ciliberto_prl} we decompose this into two terms for each particle as
\begin{equation}
\begin{gathered}
Q_{ii} = -\int_{0}^{\tau}{k_{iz} z_i\dot{z_i}dt}\\
Q_{ij} = -\int_{0}^{\tau}{\epsilon k_{jz} z_j\dot{z_i}dt}; i\neq j
\end{gathered}
\end{equation}
While the first term can be represented as a conservative force and simplifies to $Q_{ii} = -k_{iz}\left[\frac{1}{2}z_i^2\right]^\tau_0$, the second term represents the heat transferred between the two particles.

Performing work on the system of coupled particles requires varying either the conservative potential, $U_i = \frac{1}{2} k_{iz}z_i^2$ or the hydrodynamic coupling, $f_j = \epsilon k_{jz} z_j$ between them. Since $\epsilon$ is maintained constant during the engine protocols used in our experiment, both processes occur simultaneously.
To calculate the work done during an isothermal process where the parameters $k_{1z}$ and $k_{2z}$ are linearly increased/decreased, we used the standard definitions of Stochastic thermodynamics. Work done, $W_{i}$ can be decomposed into two terms as
\begin{equation}
\begin{gathered}
W_i = W_{ii} + W_{ij}; i\neq j\\
W_{ii} = \int_{0}^{\tau}{\frac{\partial U_i}{\partial t}dt} = \int_{0}^{\tau}{\frac{1}{2}\dot{k_{iz}}z_i^2dt}\\
W_{ij} = \int_{0}^{\tau}{f_j \circ\dot{z_i}dt} = \int_{0}^{\tau}{\epsilon k_{jz} z_j\dot{z_i}dt}; i\neq j
\end{gathered}
\label{work}
\end{equation}
where, $W_{ii}$s are work done by conservative optical forces and $W_{ij}$s are those by non-conservative hydrodynamic coupling. The total efficiency of the engine can be defined as 
\begin{equation}
\mathcal{E}_i = \frac{\langle W_i\rangle}{\langle Q_i\rangle}
\end{equation}
where, $W_i$ is the average work done at the end of each cycle and $Q_i$ is the heat transferred to the engine from the hot reservoir.

\subsection{Averaging considerations in calculating thermodynamic quantities}
\begin{figure}
\centering
\includegraphics[width=0.99\linewidth]{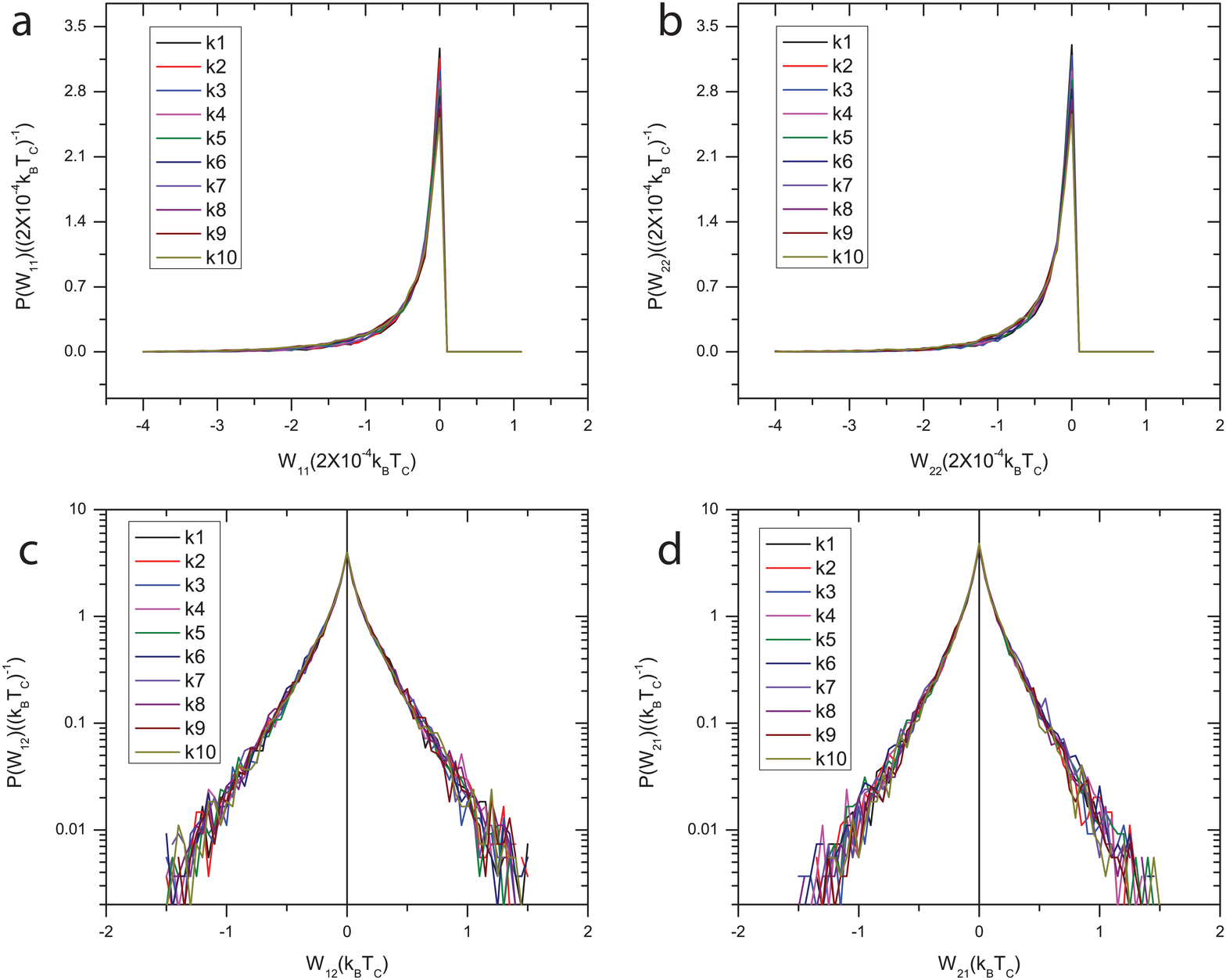}
\caption{\textbf{Work distributions along the hot isotherm.} \textbf{a - d} shows probability distributions of work done by the engine $W_{11}$, $W_{22}$, $W_{12}$ and $W_{21}$ respectively for various $k_{iz}$ along the hot isotherm. $k1\rightarrow k10$ represent $P(W_{ii})$ and $P(W_{ij})$ plotted for bins with a decreasing order in $k_{iz}$ from $k_{max}$ to $k_{min}$.}
\label {FigS4}
\end{figure}

While Eq.s \ref{work} provide a recipe to calculate the work done, the integrals should converge over a reasonable experimental duration to achieve appropriate averaging. To determine the averaging required for convergence of $W_{ii}$s and $W_{ij}$s, we observe their probability distributions along the hot isotherms calculated over a single time step of our observations = 2ms in Fig. S\ref{FigS4}. The trajectory of the particle during the isotherm was divided into 10 equal bins such that change in $k_{iz}$ was negligible across each of them. $W_{ii}$ and $W_{ij}$ were calculated using Eq.s \ref{work} over each time step of measurement and probability distributions, $P(W_{ii})$ and $P(W_{ij})$ corresponding to each bin were plotted independently. Fig.s S\ref{FigS4} a and b show such distributions of $W_{ii}$s, where $k1\rightarrow k10$ represent $P(W_{ii})$ plotted for bins with a decreasing order in $k_{iz}$ from $k_{max}$ to $k_{min}$. $W_{ii}<0$ as $z_i^2 >0$ and $\dot{k_{iz}}<0$ during isothermal expansion and indicates work done by the system on the surroundings. The summation of these $W_{ii}$s over the isotherm sampled $n = 3500$ times in our experiments converges as in the case of isolated engines to $n\langle W_{ii}\rangle<0$ \cite{Bechinger_Nature}. Similar distributions for $W_{ij}$s (Fig. S\ref{FigS4} c and d), however, are distributed equally about $W_{ij} = 0$ with a standard deviation of $\sigma(W_{ij})= 0.5 k_BT_C$. By central limit theorem, the probability distribution of the summation over n terms along the isotherm would have a standard deviation $0.5 \sqrt{n} k_BT_C$ and a mean 0. Thus, $\langle W_{ij}\rangle = 0$ if the isotherm is executed a sufficiently large number of times and averaged. Over finite experimental durations, however, the error in measuring $\langle W_{ij}\rangle$ could be significant. Reducing the error in such a measurement, the standard error of mean to even $10\%$ of $\sigma(W_{ij})$ requires us to average over $100Xn = 350,000$ cycles. Such large number of cycles are impractical in our experiments. In our analysis, we assumed $\langle W_{ij}\rangle = 0$, the average expected from central limit theorem.

\subsection{Active temperature}
Given  $\langle W_{ij}\rangle = 0$, we could define an active temperature as in previous studies. The total work done over the Stirling cycle, $W$ can be written as summations over the hot and cold isotherms that are performed at the same rate in out experiment i.e. $\dot{k_z}(hot)= -\dot{k_z}(cold)$
\begin{equation}
\begin{gathered}
W = \sum_{hot}{W_{ii}}+\sum_{cold}{W_{ii}}\\
= \frac{1}{2\nu}\dot{k}(hot)\left[\sum_{hot}z_i^2-\sum_{cold}z_i^2 \right]
\end{gathered}
\end{equation}
where $\nu$ is the sampling frequency = 500Hz in our experiment. Using equipartition theorem, we could define active temperature as $T_{iact} = \frac{k_z\langle z_i^2\rangle}{k_B}$. Average work done over the Stirling cycle can then be written as
\begin{equation}
\langle W \rangle = \frac{1}{2\nu k_B}\dot{k}(hot)\left[\sum_{hot}\frac{T_{iact}(hot)-T_{iact}(cold)}{k_{iz}}\right]
\end{equation}
Thus, average work done over the cycle would be proportional to the difference in $T_{act}$ between the hot and cold isotherms. In our experiments in Fig. 5 of the main paper, we observed that this difference increases due to the mutual interactions and leads to better performance.

Finally, since we did not observe any changes in the probability distributions along x and y axes as discussed in Fig. S3, the above analysis can be used to calculate performance along these directions assuming $\alpha_{i} = 0$.

\section{Brownian dynamics simulations}
In our analysis in Fig. 3f of the main paper we simulated the equations of motion Eq. \ref{eq_motion1} through Brownian dynamics algorithms. Particle trajectories were computer generated using conventional Brownian dynamics algorithm \cite{branka_algorithms, branka_algorithms1} for a time step $dt = 0.2ms$. We observed that $dt$ was sufficiently small that more sophisticated algorithms such as stochastic Runge-Kutta algorithm, stochastic expansion etc. \cite{branka_algorithms, branka_algorithms1} resulted in changes only in the fourth significant figure.

\begin{figure}
\centering
\includegraphics[width=0.99\linewidth]{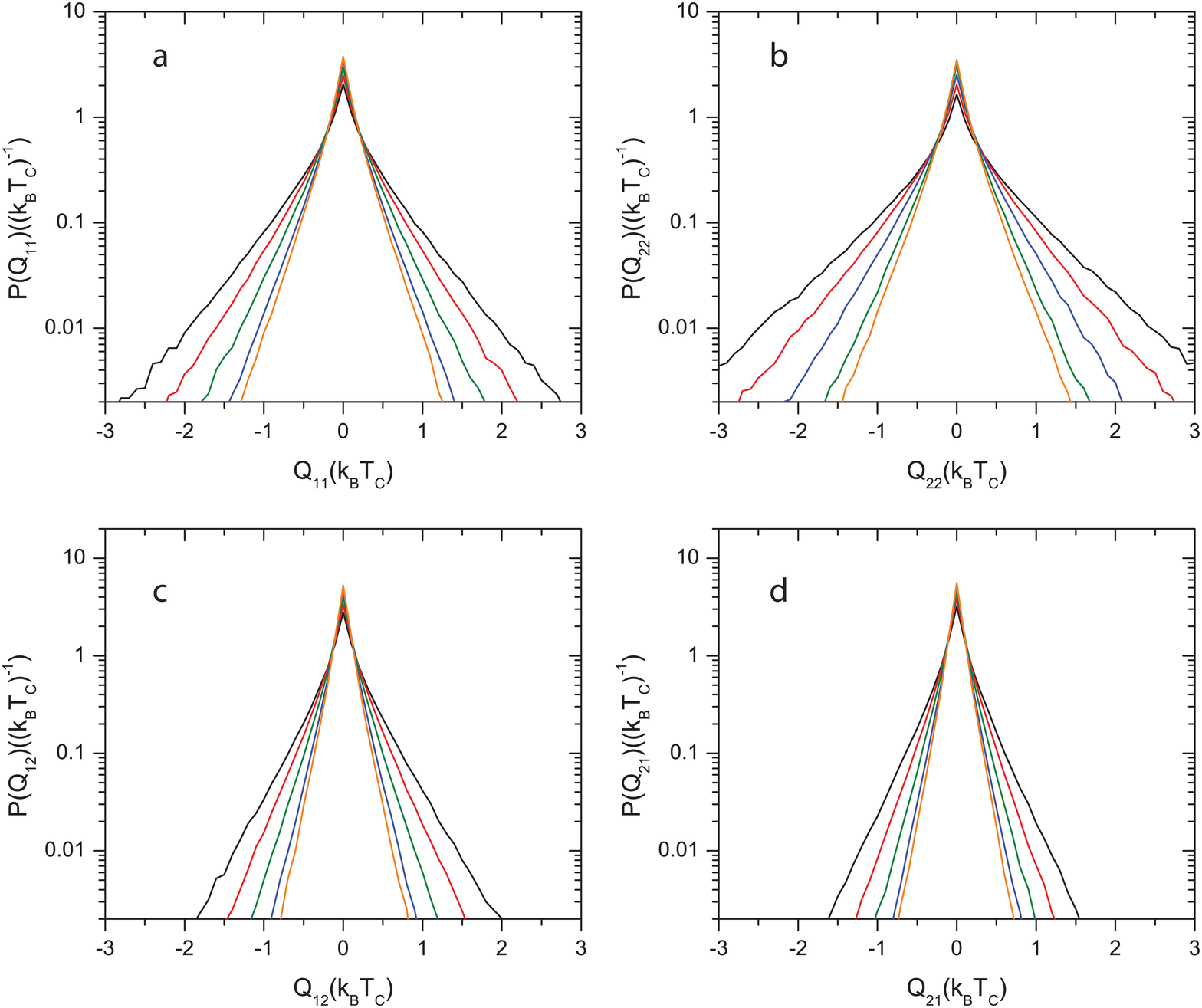}
\caption{\textbf{Simulated heat flows at various} $\mathbf{\left| \alpha_i \right| }$ \textbf{at constant} $\mathbf{\epsilon}$. \textbf{a - d} shows probability distributions of heat fluxes $Q_{11}$, $Q_{22}$, $Q_{12}$ and $Q_{21}$ respectively for the noise intensities $\left| \alpha_1 \right| / \sqrt{2\gamma k_{B}T_C} = 1$ (orange), $5$ (blue), $10$ (green), $15$ (red) and $20$ (black) at constant $\epsilon = 0.5$. }
\label {FigS5}
\end{figure}

\begin{figure}
\centering
\includegraphics[width=0.99\linewidth]{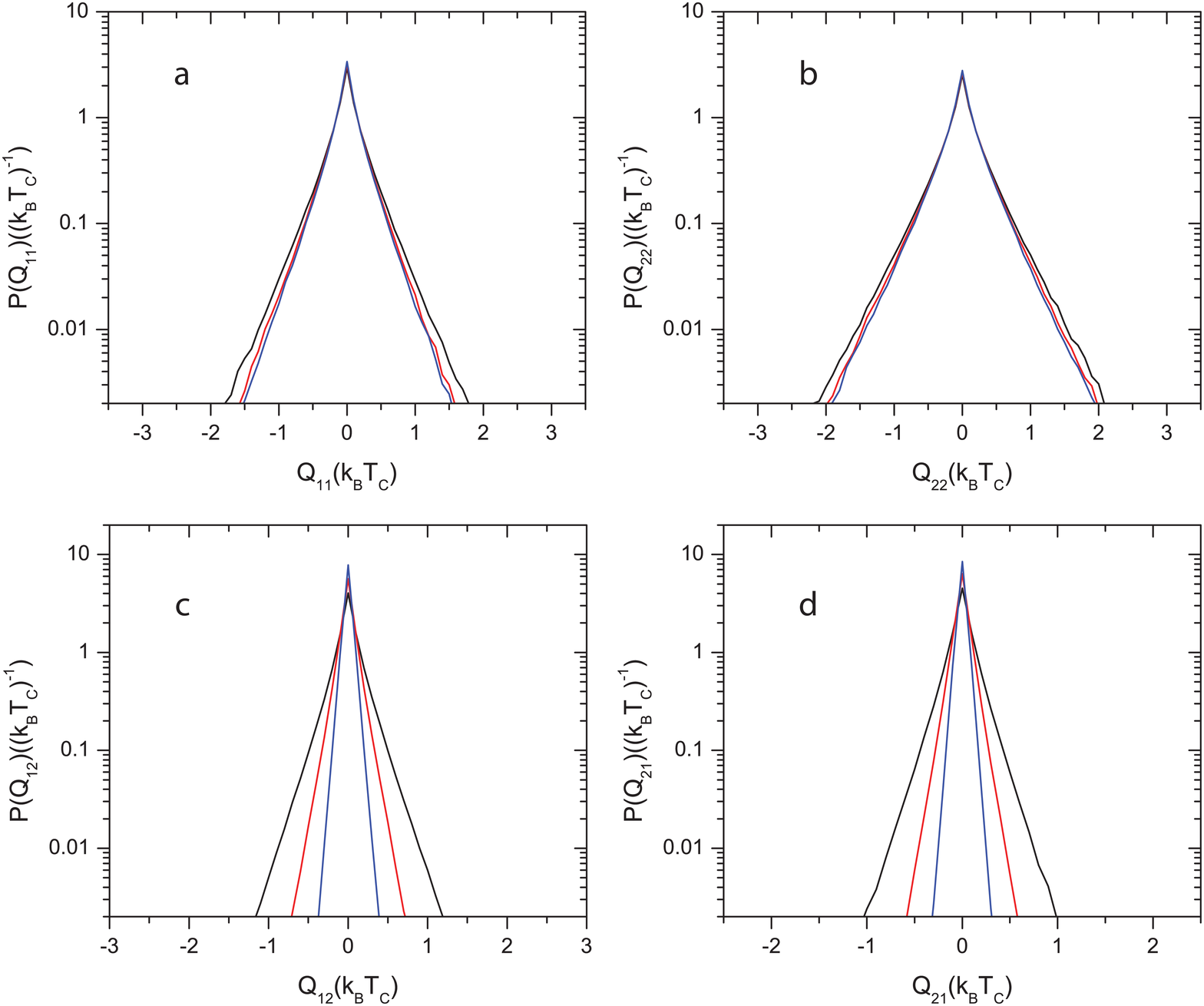}
\caption{\textbf{Simulated heat flows at various} $\mathbf{\epsilon}$ \textbf{at constant} $\mathbf{\left| \alpha_i \right| }$. \textbf{a - d} shows probability distributions of heat fluxes $Q_{11}$, $Q_{22}$, $Q_{12}$ and $Q_{21}$ respectively for various $\epsilon = 0.5$ (black), $0.3$ (red) and $0.15$ (blue) for noise intensity $\left| \alpha_1 \right| / \sqrt{2\gamma k_{B}T_C} = 10$.}
\label {FigS6}
\end{figure}

In our discussions in the main paper, we noted that unlike experiments, the simulations allowed us to independently vary $\left|\alpha_i\right|$ and $\epsilon$ and this can be leveraged to obtain insights into the nature of the interactions. To this extent, we varied the magnitude of the noise, $\left|\alpha_i\right|$ while $\epsilon$ was maintained constant and observed the corresponding variations in heat currents in Fig. S\ref{FigS5}. The results in Fig. S\ref{FigS5} closely match the experimental counterparts in Fig.s 3 a-d of the main paper. Similar observations made by alternately fixing $\left|\alpha_i\right|$ as $\epsilon$ was varied are presented in Fig. S\ref{FigS6}. In contrast to Fig. S\ref{FigS5}, we observe that the changes in stochastic heat currents in Fig. S\ref{FigS6} are relatively small. However, the experimental data in which these two cannot be independently varied did show a significant increase in standard deviations in heat transfer (Fig. 3 a-d of the main paper). Thus, while arranging the optical traps at close separation does increase the hydrodynamic coupling between them, the increase in mutual interactions arise largely from the associated increase in Brownian vortexes.

\begin{figure}
\centering
\includegraphics[width=0.99\linewidth]{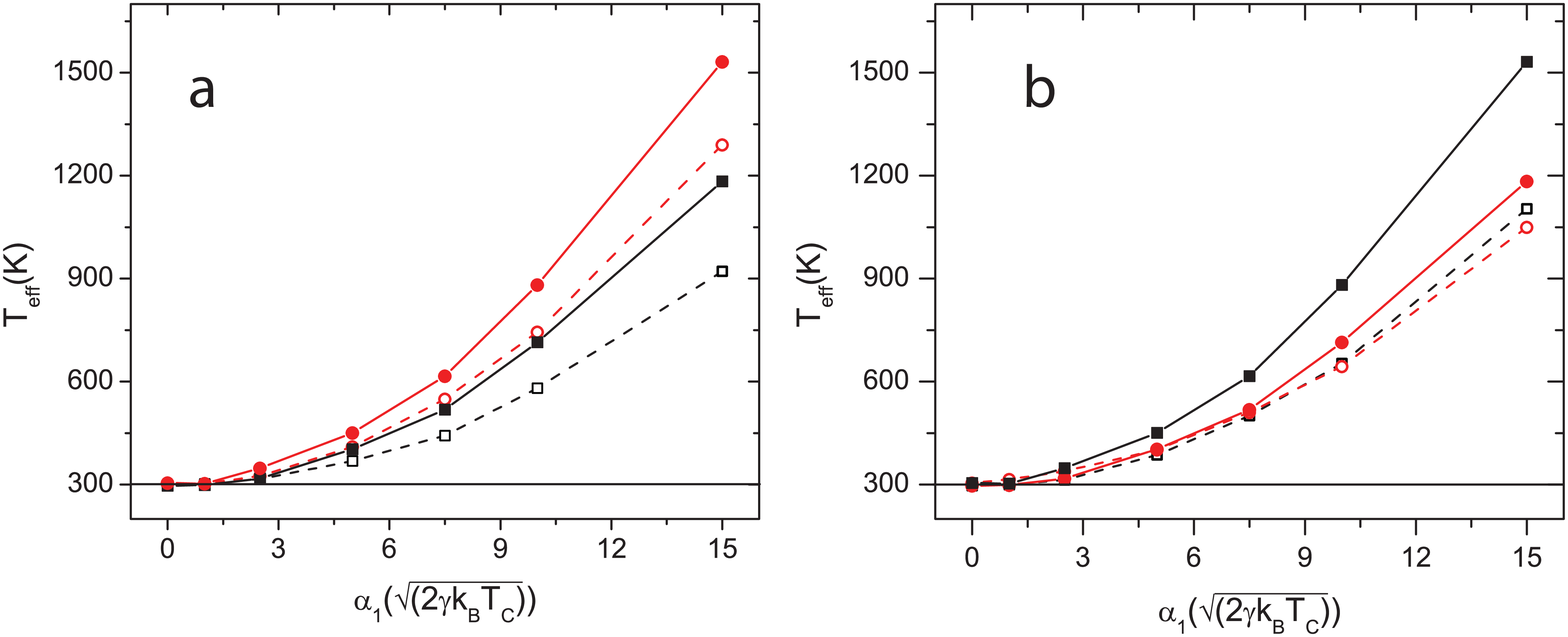}
\caption{\textbf{Adjusting $T_{eff}$ with simulation parameters.} \textbf{a} shows variations in $T_{1eff}$ (red circles) and $T_{2eff}$ (black squares) with $\left| \alpha_1 \right|$ for $\epsilon = 0.5$ (closed symbols) and $0.15$ (open symbols). \textbf{b} shows a similar plot for various ratios of noise intensities $\langle\left| \alpha_2\right| \rangle /\langle\left| \alpha_1\right| \rangle = 1.33$ (closed symbols) and $1$ (open symbols).}
\label {FigS7}
\end{figure}

Finally, in Fig. S\ref{FigS7} we explore the changes in heat flow as we vary the simulation parameters. All deterministic parameters were maintained at their approximate experimental values i.e. $k_{1z} = 0.49 pN/\mu m$, $k_{2z} = 0.66pN/\mu m$, $T = 300K$, $dt = 0.2ms$, $\mu = 0.8 mPa.s$, $\sigma = 2 \mu m$, $\langle\left| \alpha_2\right| \rangle/\langle\left| \alpha_1\right| \rangle =1.33$. To quantify the interactions, we used the active temperature $T_{iact}$ defined in the previous section. $T_{iact}$ was observed to increase with $\left|\alpha_1\right|$ for both $\epsilon = 0.5$ and $0.15$ (Fig. S7), but it was significantly higher at close separation ($\epsilon = 0.5$). In our experiments, the strength of interactions $\left|\alpha_{i}\right|$ changes with $k_{1z}$ and $k_{2z}$ during the isotherms, this could lead to different $T_{act}$ along them as observed in Fig. 5 of the main paper. Further, varying $\langle\left| \alpha_2\right| \rangle/\langle\left| \alpha_1\right| \rangle$ which we had fixed at 1.33 could also allow us to adjust the model to fit the experimental data. In Fig. S\ref{FigS7} b, we show that by changing it to 1, we could even reach a scenario where $T_{1act} = T_{2act}$. In conclusion, while the heat flows arising from interactions due to non-conservative flows can be captured by our model, it does not predict relations between $T_{iact}$ and $k_{iz}$. Arriving at such relations continues to remain an open question after our simulations. Nevertheless, the model can be appropriately adjusted with parameters $\left|\alpha_1\right|$ and $\langle\left| \alpha_2\right| \rangle/\langle\left| \alpha_1\right| \rangle$ to capture the heat flows and understand the increase in performance.

\bibliography{paper_ref}

\newpage

\newpage

\newpage
	
\newpage

\newpage

\newpage

\newpage
